\documentclass[sigconf]{acmart}

\AtBeginDocument{%
  \providecommand\BibTeX{{%
    \normalfont B\kern-0.5em{\scshape i\kern-0.25em b}\kern-0.8em\TeX}}}

\usepackage{balance}  
\usepackage{stmaryrd}
\usepackage{booktabs} 
\usepackage{framed}

\usepackage[linesnumbered,boxed,ruled]{algorithm2e}
\usepackage{enumitem}
\usepackage{dsfont}
\usepackage{graphicx, subfigure}

\newcommand{\ie}{{\em i.e., \xspace}}
\newcommand{\nosection}[1]{\vspace{2pt}\noindent\textbf{#1.}}
\newcommand{\modelname}{\texttt{CAESAR}}
\newcommand{\companyname}{Ant}
\newenvironment{protocol}[1][htb]
  {
   \begin{algorithm}[#1]%
  }{\end{algorithm}}
  
\setcopyright{acmcopyright}
\copyrightyear{2021}
\acmYear{2021}
\setcopyright{acmcopyright}\acmConference[KDD '21]{Proceedings of the 27th ACM SIGKDD Conference on Knowledge Discovery and Data Mining}{August 14--18, 2021}{Virtual Event, Singapore}
\acmBooktitle{Proceedings of the 27th ACM SIGKDD Conference on Knowledge Discovery and Data Mining (KDD '21), August 14--18, 2021, Virtual Event, Singapore}
\acmPrice{15.00}
\acmDOI{10.1145/3447548.3467210}
\acmISBN{978-1-4503-8332-5/21/08}

\begin{document}
\title{When Homomorphic Encryption Marries Secret Sharing: \\Secure Large-Scale Sparse Logistic Regression and Applications in Risk Control}
\titlenote{Jun Zhou is the corresponding author.}
\fancyhead{}
\author{Chaochao Chen$^{1}$, Jun Zhou$^{1}$, Li Wang$^{1}$, Xibin Wu$^{1}$, Wenjing Fang$^{1}$, Jin Tan$^{1}$, Lei Wang$^{1}$, Alex X. Liu$^{1}$, Hao Wang$^{2}$, Cheng Hong$^{3}$}
\affiliation{
  $^{1}$Ant Group, $^{2}$Shandong Normal University, $^{3}$Alibaba Group\\
\{chaochao.ccc, jun.zhoujun, raymond.wangl, xibin.wxb, bean.fwj, tanjin.tj, shensi.wl, alexliu\}@antgroup.com \\
wanghao@sdnu.edu.cn, vince.hc@alibaba-inc.com
}


\renewcommand{\shortauthors}{Chaochao Chen et al.}


\begin{abstract}
Logistic Regression (LR) is the most widely used machine learning model in industry for its efficiency, robustness, and interpretability. 
Due to the problem of data isolation and the requirement of high model performance, many applications in industry call for building a secure and efficient LR model for multiple parties.
Most existing work uses either Homomorphic Encryption (HE) or Secret Sharing (SS) to build secure LR. 
HE based methods can deal with high-dimensional sparse features, but they incur potential security risks.
SS based methods have provable security, but they have efficiency issue under high-dimensional sparse features. 
In this paper, we first present \modelname, which combines HE and SS to build secure large-scale sparse logistic regression model and achieves both efficiency and security. 
We then present the distributed implementation of \modelname~for scalability requirement.
We have deployed \modelname~in a risk control task and conducted comprehensive experiments.
Our experimental results show that \modelname~improves the state-of-the-art model by around 130 times.
\end{abstract}


\maketitle

\section{Introduction}\label{intro}

Logistic Regression (LR) and other machine learning models have been popularly deployed in various applications by different kinds of companies, e.g., advertisement in e-commerce companies \cite{zhou2017kunpeng}, disease detection in hospitals \cite{chen2017machine}, 
and fraud detection in financial companies \cite{zhang2019distributed}. 
In reality, there are also increasingly potential gains if different organizations could collaboratively combine their data for data mining and machine learning. 
For example, health data from different hospitals can be used together to facilitate more accurate diagnosis, while financial companies can collaborate to train more effective fraud-detection engines.
Unfortunately, this cannot be done in practice due to competition and regulation reasons. 
That is, data are isolated by different parties, which is also known as the ``isolated data island'' problem \cite{yang2019federated}.

To solve this problem, the concept of privacy-preserving or secure machine learning is introduced. Its main goal is to combine data from multiple parties to improve model performance while  protecting data holders from any possibility of information disclosure. The security and privacy concerns, together with the desire of data combination, pose an important challenge for both academia and industry. 
To date, 
Homomorphic Encryption (HE) \cite{paillier1999public} and secure Multi-Party Computation (MPC) \cite{yao1982protocols} are two popularly used techniques to solve the above challenge. 
For example, many HE-based methods have been proposed to train LR and other machine learning models using participants' encrypted data, as a centralized manner \cite{yuan2013privacy,gilad2016cryptonets,badawi2018alexnet,hesamifard2017cryptodl,xu2019cryptonn,aono2016scalable,kim2018secure,chen2018logistic,Han2019logistic,esperancca2017encrypted,jiang2018securelr}. 
Although these data are encrypted, they can be abused by the centralized data holder, which raises potential information leakage risk.
There are also several HE-based methods which build machine learning models in a decentralized manner \cite{xie2016privlogit,hardy2017private}, i.e., data are still held by participants during model training procedure. 
These methods process features by participants themselves and therefore can handle high-dimensional sparse features. 
Moreover, participants only need to communicate encrypted reduced information during model training, and therefore the communication cost is low. 
However, models are exposed as plaintext to server or participants during each iteration in the training procedure, which can be used to infer extra private information \cite{li2019quantification}, even in semi-honest security setting. 
In a nutshell, these models are non-provably secure. 

\begin{figure}[t]
\centering 
\includegraphics[width=7cm]{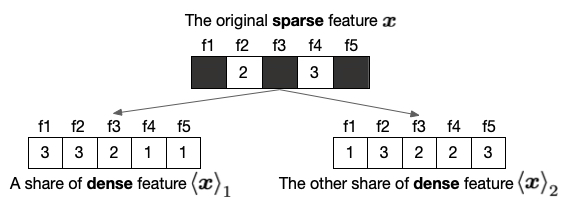} 
\vskip -0.15in
\caption{Sparse feature becomes dense features after using secret sharing in $\mathds{Z}_4$. }
\label{fig:example}
\vskip -0.16in
\end{figure}

Besides HE-based methods, in literature, many MPC---mostly Secret Sharing (SS) \cite{shamir1979share}---based protocols, are proposed for secure machine learning models, e.g., linear regression \cite{gascon2017privacy}, logistic regression \cite{shi2016secure,chen2020secret}, tree based model \cite{fang2020hybrid}, neural network \cite{rouhani2018deepsecure,wagh2018securenn,zheng2020industrial}, recommender systems \cite{chen2020secure}, and general machine learning algorithms \cite{mohassel2017secureml,demmler2015aby,mohassel2018aby,li2018privpy}. 
Besides academic research, large companies are also devoted to develop secure and privacy preserving machine learning systems. 
For example, Facebook opensourced CrypTen\footnote{https://github.com/facebookresearch/CrypTen/} 
and Visa developed ABY$^3$ \cite{mohassel2018aby}. 
Comparing with HE based methods, SS based methods have provable security. 
However, since their idea is secretly sharing the participants' data (both features and labels) on several servers for secure computation, the features become dense after using secret sharing even they are sparse beforehand. 
Figure \ref{fig:example} shows an example, where the original feature is a 5-dimensional sparse vector. While after secretly share it to two parties, it becomes two dense vectors. 
Therefore, they cannot handle sparse features efficiently and have high communication complexity when dataset gets large, as we will analyze in Section \ref{sec-model-adv}. 

\subsection{Gaps Between Research and Practice}
We analyze the gaps between research and practice from the following three aspects. 
(1) \textit{Number of participants}. 
In research, most existing works usually assume that any number of parties can jointly build privacy-preserving machine learning models \cite{mohassel2017secureml,hardy2017private,demmler2015aby,mohassel2018aby}, and few of them are customized to only two-party setting. 
However, in practice, it is always two parties, as we will introduce in Section \ref{sec-exp}. 
This is because it usually involves more commercial interests and regulations when there are three or more parties. 
(2) \textit{Data partition}. Many existing researches are suitable for both data vertically partition and data horizontally partition settings. In practice, when participants are large business companies, they usually have the same batch of samples but different features, i.e., data vertically partition. There are limited researches focus on this setting and leverage its characteristic. 
(3) \textit{Feature sparsity}. Existing studies usually ignore that features are usually high-dimensional and sparse in practice, which is usually caused by missing feature values or feature engineering such as one-hot. 
Therefore, how to build secure large-scale LR under data vertically partition is a challenging and valuable task for industrial applications. 

\subsection{Our Contributions}

\nosection{New protocols by marrying HE and SS} 
We present \modelname, a seCure lArge-scalE SpArse logistic Regression model by marrying HE and SS.  
To guarantee security, we secretly share model parameters to both parties, rather than reveal them during model training \cite{xie2016privlogit,hardy2017private}. 
To handle large-scale sparse data when calculating predictions and gradients, we propose a secure sparse matrix multiplication protocol based on HE and SS, which is the key to scalable secure LR. 
By combining HE and SS, \modelname~has the advantages of both efficiency and provable security. 

\nosection{Distributed implementation} 
Our designed implementation framework is comprised of a coordinator and two distributed clusters on both participants' sides. 
The \textit{coordinator} controls the start and terminal of the clusters. 
In each \textit{cluster}, 
we borrow the idea of parameter server \cite{li2014scaling,zhou2017kunpeng} and distribute data and model on \textit{servers} who learn model parameters following our proposed \modelname. 
Meanwhile, each server delegates the most time-consuming encryption operations to multiple \textit{workers} for distributed encryption. By vertically distribute data and model on servers, and delegate encryption operations on workers, our implementation can scale to large-scale dataset.

\nosection{Real-world deployment and applications} 
We deployed \modelname~in a risk control task in \companyname~Financial (\companyname, for short), and conducted comprehensive experiments. 
The results show that \modelname~significantly outperforms the state-of-the-art secure LR model, i.e., SecureML, especially under the situation where network bandwidth is the bottleneck, e.g., limited communication ability between participants or high-dimensional sparse features. 
Taking our real-world risk control dataset, which has around 1M samples and 100K features (3:7 vertically partitioned), as an example, it takes 7.72 hours for \modelname~to finish an epoch when bandwidth is 10Mbps and batch size is 4,096. In contrast, SecureML needs to take 1,005 hours under the same setting---a \textbf{130x speedup}. 
To the best of our knowledge, \modelname~is the first secure LR model that can handle such large-scale datasets efficiently. 

\section{Related Work}\label{background}
In this section, we review literatures on Homomorphic Encryption (HE) based privacy-preserving Machine Learning (ML) and Multi-Party Computation (MPC) based secure ML. 

\subsection{HE based Privacy-Preserving ML}\label{bg-he}
Most existing HE based privacy-preserving ML models belong to centralized modelling. 
That is, private data are first encrypted by participants and then outsourced to a server who trains ML models as a centralized manner using HE techniques. 
Various privacy-preserving ML models are built under this setting, including least square \cite{esperancca2017encrypted}, logistic regression \cite{aono2016scalable,kim2018secure,chen2018logistic,Han2019logistic}, and neural network \cite{yuan2013privacy,gilad2016cryptonets,badawi2018alexnet,hesamifard2017cryptodl,xu2019cryptonn}. 
However, this kind of approach suffers from data abuse problem, since the server can do whatever computations with these encrypted data in hand. 
The data abuse problem may further raise potential risk of data leakage. 

There are also some researches focus on training privacy-preserving ML models using HE under a decentralized manner. 
That is, the private data are still held by participants during model training. 
For example, Wu et al. proposed a secure logistic regression model for two parties, assuming that one party has features and the other party has labels \cite{wu2013privacy}. 
Other researches proposed to build linear regression \cite{hall2011secure} and logistic regression \cite{xie2016privlogit} under horizontally partitioned data. 
As we described in Section \ref{intro}, features partitioned vertically is the most common setting in practice. 
Thus, the above methods are difficult to apply into real-world applications. 
The most similar work to ours is vertically Federated Logistic Regression (FLR) \cite{hardy2017private}. 
FLR trains logistic regression in a decentralized manner, i.e., both private features and labels are held by participants during model training. 
Besides, participants only need to communicate compressed encrypted information with each other and a third-party server, thus its communication cost is low. 
However, it assumes there is a third-party that does not collude with any participants, which may not exist in practice. 
Moreover, model parameters are revealed to server or participants in plaintext during each iteration in the training procedure, which can be used to infer extra information and cause information leakage in semi-honest security setting \cite{li2019quantification}. 
In contrast, in this paper, we propose to marry HE and SS to build secure logistic regression. 


\subsection{MPC based Secure ML}\label{sec-related-mpc}
Besides HE, MPC is also popularly used to build secure ML systems in literature. 
First of all, there are some general-purpose MPC protocols such as VIFF \cite{damgaard2009asynchronous} and SPDZ \cite{damgaard2012multiparty} that can be used to build secure logistic regression model \cite{chen2019secure}. 
Secondly, there are also MPC protocols for specific ML algorithms, e.g., garbled circuit and HE based neural network \cite{juvekar2018gazelle}, garbled circuit based linear regression \cite{gascon2017privacy}, logistic regression \cite{shi2016secure}, and  neural network \cite{rouhani2018deepsecure,agrawal2019quotient}. 

In recent years, there is a trend of building general ML systems using Secret Sharing (SS). 
For example, Demmler et al. proposed ABY \cite{demmler2015aby}, which combines Arithmetic sharing (A), Boolean sharing (B), and Yao's sharing (Y) for general ML. Mohassel and Zhang proposed SecureML \cite{mohassel2017secureml}, which optimized ABY for vectorized scenario so as to compute multiplication of shared matrices and vectors. 
Later on, Mohassel and Rindal proposed ABY$^3$ \cite{mohassel2018aby}, which extended ABY and SecureML to a three-server mode setting and thus has better efficiency. 
Li et al. proposed PrivPy \cite{li2018privpy} that works under four-server mode. 
The above approaches are provably secure and their basic idea is secretly sharing the data/model among multi-parties/servers. 
Under thus circumstances, these systems cannot scale to high-dimensional data even when these data are sparse, 
since secret sharing will make the sparse data dense. 

Recently, Phillipp et al. proposed ROOM \cite{schoppmann2019make} to solve the data sparsity problem in machine learning. However, it needs to reveal data sparseness which may cause potential information leakage. For example, when a dataset contains only binary features (0 or 1), a dense sample directly reflects that its features are all ones if using ROOM. 
Besides, when it involves matrix multiplication, ROOM still needs cryptographical techniques to generate Beaver triples \cite{beaver1991efficient}, which limits its efficiency for training. 
In this paper, we propose to combine HE and SS to improve the communication efficiency of the existing MPC based secure logistic regression, which can not only protect data sparseness, but also avoid the time-consuming Beaver triples generation procedure. 

\section{Preliminaries}	
In this section, we briefly describe the setting and threat model of our proposal, and present some background knowledge.

\subsection{Data Vertically Partitioned by Two-Parties}\label{pre-setting}

In this work, we consider secure protocols for two parties who want to build secure logistic regression  together. 
Moreover, existing works on secure logistic regression mainly focus on two cases based on how data are partitioned between participants, i.e., \textit{horizontally data partitioning} that denotes each party has a subset of the samples with the same features, and \textit{vertically data partitioning} which means each party has the same samples but different features \cite{hall2011secure}. 
In this paper, we focus on vertically partitioning setting, since it is more common in industry. 
It becomes more general as long as one of the participants is a large company who has hundreds of millions of customers. 

Note that, in practice, when participants collaboratively build secure logistic regression under vertically data partitioning setting, the first step is matching sample IDs between participants. 
\textit{Private Set Intersection} technique \cite{pinkas2014faster} is commonly used to get matched IDs privately. 
We omit its details in this paper and only focus on the machine learning part. 

\subsection{Threat Model}
We consider the standard \emph{semi-honest model}, the same as the existing methods \cite{mohassel2017secureml,li2018privpy}, where a probabilistic polynomial-time adversary with semi-honest behaviors is considered. In this security model, the adversary may corrupt and control one party (referred as to {\em the corrupted party}), and try to obtain information about the input of the other party (referred as to {\em the honest party}). During the protocol executed by the honest party, the adversary will follow the protocol specifically, but may attempt to obtain additional information about the honest party's input by analyzing the corrupted party's \emph{view}, i.e., the transcripts it receives during the protocol execution. 
The detailed definition of the semi-honest security model can be found in Appendix \ref{appen-a}. 

\subsection{Additive Secret Sharing}\label{sec-pre-ss}

We use the classic additive secret sharing scheme under our two party ($\mathcal{A}$ and $\mathcal{B}$) setting \cite{shamir1979share,boyle2015function}. 
Let $\phi = 2 ^{l}$ be a large integer, $x$ and $y$ be non-negative integers and $0 < x, y \ll \phi$, and $\mathds{Z}_\phi$ be the group of integers module $\phi$. 
Assuming that $\mathcal{A}$ wants to \textbf{share} a secret $x$ with $\mathcal{B}$, $\mathcal{A}$ first randomly samples an integer $r$ in $\mathds{Z}_\phi$ as a share $\left\langle x \right\rangle_2$ and sends it to $\mathcal{B}$, and then calculates $x - r ~\text{mod}~ \phi$ as the other share $\left\langle x \right\rangle_1$ and keeps it itself. 
To this end, $\mathcal{A}$ has $\left\langle x \right\rangle_1$ and $\mathcal{B}$ has $\left\langle x \right\rangle_2$ such that $\left\langle x \right\rangle_1$ and $\left\langle x \right\rangle_2$ are randomly distributed integers in $\mathds{Z}_\phi$ and $x = \left\langle x \right\rangle_1 + \left\langle x \right\rangle_2 ~\text{mod}~ \phi$. 
Similarlly, assume $\mathcal{B}$ has a secret $y$ and after shares it with $\mathcal{A}$, $\mathcal{A}$ has $\left\langle y \right\rangle_1$ and $\mathcal{B}$ has $\left\langle y \right\rangle_2$. 

\nosection{Addition}
Suppose $\mathcal{A}$ and $\mathcal{B}$ want to secretly calculate $x + y$ in secret sharing, $\mathcal{A}$ computes $\left\langle z \right\rangle_1 = \left\langle x \right\rangle_1 + \left\langle y \right\rangle_1 ~\text{mod}~ \phi$ and $\mathcal{B}$ computes $\left\langle z \right\rangle_2 = \left\langle x \right\rangle_2 + \left\langle y \right\rangle_2 ~\text{mod}~ \phi$ and each of them gets a share of the addition result. 
To \textbf{reconstruct} a secret, one party just needs to send its share to the other party, and then reconstruct can be done by $z = \left\langle z \right\rangle_1 + \left\langle z \right\rangle_2 ~\text{mod}~ \phi$. 

\nosection{Multiplication}
Most existing secret sharing multiplication protocol are based on Beaver's triplet technique \cite{beaver1991efficient}. 
Specifically, to multiply two secretly shared values $\langle x \rangle$ and $\langle y \rangle$ between two parties, they need a shared triple (Beaver's triplet) $\langle u \rangle$, $\langle v \rangle$, and $\langle w \rangle$, where $u, v$ are uniformly random values in $\mathds{Z}_\phi$ and $w=u \cdot v$ mod $\phi$. 
They then make communication and local computations, and finally each of the two parties gets $\langle z \rangle_1$ and $\langle z \rangle_2$, respectively, such that $\langle x \rangle \cdot \langle y \rangle =\langle z \rangle_1+\langle z \rangle_2$. 

\nosection{Supporting real numbers and vectors}
We use fix-point representation to map real numbers to $\mathds{Z}_\phi$ \cite{mohassel2017secureml,li2018privpy}. 
Assume $x \in [-p, p]$ is a real number where $0 < p \ll \phi /2$, it can be represented by $\lfloor 10^c x \rfloor$ if $x \ge 0$ and $\lfloor 10^c x \rfloor + \phi$ if $x < 0$, 
where $c$ determines the precision of the represented real number, i.e., the fractional part has $c$ bits at most. 
After this, it can be easily vectorized to support matrices, e.g., SS based secure matrix multiplication in \cite{de2017efficient}. 

\subsection{Additive Homomorphic Encryption}
Additive HE methods, e.g., Okamoto-Uchiyama encryption (OU) \cite{okamoto1998new} and Paillier \cite{paillier1999public}, are popularly used in machine learning algorithms \cite{aono2016scalable}, as described in Section \ref{bg-he}. 
The use of additive HE mainly has the following steps \cite{acar2018survey}: 
\begin{itemize}[leftmargin=*] \setlength{\itemsep}{-\itemsep}
    \item \textbf{Key generation. } One participant generates the public and secret key pair $(pk, sk)$ and publicly distributes $pk$ to the other participant.
    \item \textbf{Encryption. } Given a plaintext $x$, it is encrypted using $pk$ and a random $r$, i.e., $\llbracket x \rrbracket=\textbf{Enc}(pk; x, r)$, where $\llbracket x \rrbracket$ denotes the ciphertext and $r$ makes sure the ciphertexts are different in multiple encryptions even when the plaintexts are the same. 
    \item \textbf{Homomorphic operation. } Given two plaintexts ($x$ and $y$) and their corresponding ciphertexts ($\llbracket x \rrbracket$ and $\llbracket y \rrbracket$), there are three types of operations for additive HE, i.e., \textbf{OP1: }$\llbracket x+y \rrbracket =x+\llbracket y \rrbracket$, \textbf{OP2: }$\llbracket x+y \rrbracket=\llbracket x \rrbracket+\llbracket y \rrbracket$, and \textbf{OP3: }$\llbracket x \cdot y \rrbracket = x \cdot \llbracket y \rrbracket$. Note that we overload `+' as the homomorphic addition operation. 
    \item \textbf{Decryption. } Given a ciphertext $\llbracket x \rrbracket$, it is decrypted using $sk$, i.e., $x=\textbf{Dec}(sk; \llbracket x \rrbracket)$. 
\end{itemize}

Similar as secret sharing, the above additive HE only works on a finite field. 
One can use similar fix-point representation approach to support real numbers in a group of integers module $\psi$, i.e., $\mathds{Z}_\psi$. 
Additive HE operations on matrix work similarly \cite{hardy2017private}.

\subsection{Logistic Regression Overview}\label{pre-ml}
We  briefly describe the key components of logistic regression as follows. 

\nosection{Model and loss}
Let $\mathcal{D}=\{(\textbf{x}_i, y_i)\}_{i=1}^n$ be the training dataset, where $n$ is the sample size, $\textbf{x}_i \in \mathds{R}^{1 \times d}$ is the feature of $i$-th sample with $d$ denoting the feature size, and $y_i$ is its corresponding label. 
Logistic regression aims to learn a model $\textbf{w}$ so as to minimize the loss $\mathcal{L}=\sum_{i=1}^n l(y_i, \hat{y_i})$, where $l(y_i, \hat{y_i})= - y \cdot log(\hat{y_i}) - (1-y) \cdot log(1 - \hat{y_i})$ with $\hat{y_i} = 1 / (1 + e ^ {-\textbf{x}_i \cdot \textbf{w}})$. 
Note that without loss of generality, we employ bold capital letters (e.g., $\textbf{X}$) to denote matrices and use bold lowercase letters (e.g., $\textbf{w}$) to indicate vectors. 

\nosection{Mini-batch gradient descent}
The logistic regression model can be learnt efficiently by minimizing the loss using mini-batch gradient descent. 
That is, instead of selecting one sample or all samples of training data per iteration, a batch of samples are selected and $\textbf{w}$ is updated by averaging the partial derivatives of the samples in the current batch. 
Let $\textbf{B}$ be the current batch, $|\textbf{B}|$ be the batch size, $\textbf{X}_B$ and $\textbf{Y}_B$ be the features and labels in the current batch, then the model updating can be expressed in a vectorized form: 
\begin{equation}\label{batch-update}
\textbf{w} \leftarrow \textbf{w} - \frac{\alpha}{|\textbf{B}|} \cdot \frac{\partial \mathcal{L}}{\partial \textbf{w}} ,
\end{equation}
where $\alpha$ is the learning rate that controls the moving magnitude and $\partial \mathcal{L} / \partial \textbf{w} = (\hat{\textbf{Y}_B} - \textbf{Y}_B)^T \cdot \textbf{X}_B$ is the total gradient of the loss with respect to the model in current batch, and we omit regularization terms for conciseness. 
From it, we can see that model updation involves many matrix multiplication operations. In practice, features are always high-dimensional and sparse, as we described in Section \ref{intro}. Therefore, sparse matrix multiplication is the key to large scale machine learning. 
The above mini-batch gradient descent not only benefits from fast convergence, but also enjoys good computation speed by using vectorization libraries. 

\begin{figure}[t]
\centering 
\includegraphics[width=0.65\columnwidth]{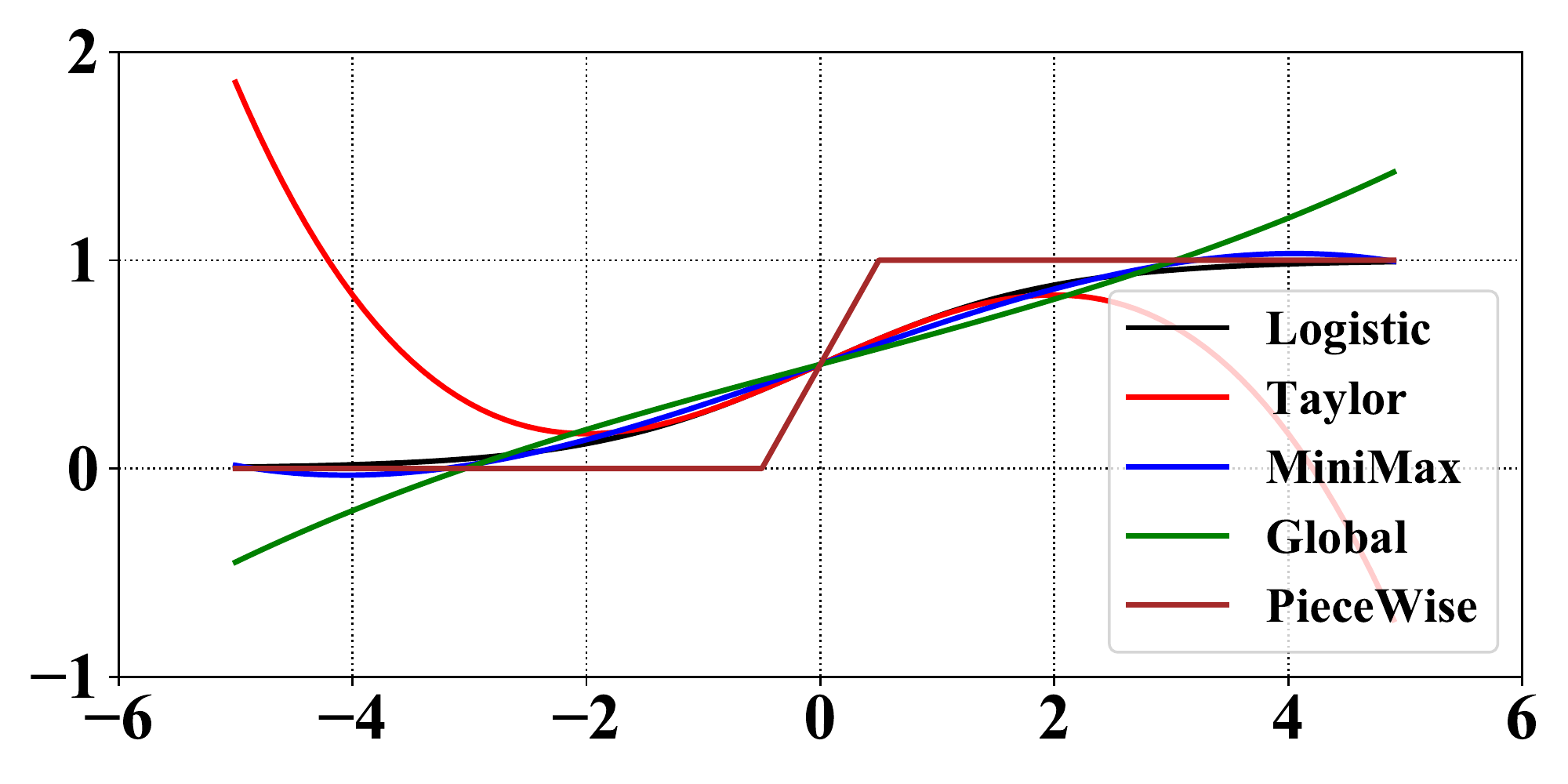} 
\vskip -0.15in
\caption{Approximation results of different methods. }
\label{fig:logistic-appro}
\vskip -0.15in
\end{figure}

\nosection{Sigmoid approximation}\label{pre-appro}
The non-linear sigmoid function in logistic regression is not cryptographically friendly. Existing researches have proposed different approximation methods for this, which include Taylor expansion \cite{hardy2017private,kim2018secure}, Minimax approximation \cite{chen2018logistic}, Global approximation \cite{kim2018secure}, and Piece-wise approximation \cite{mohassel2017secureml}. 
Figure \ref{fig:logistic-appro} shows the approximation results of these methods, where we set the polynomial degree to 3. 
In this paper, we choose the Minimax approximation method, considering its best performance. 

\section{CAESAR: Secure Large-scale Sparse Logistic Regression }	
In this section, we first describe our motivation. 
We then propose a secure sparse matrix multiplication protocol by combining HE and SS. 
Finally, we present the secure large-scale logistic regression algorithm
and its advantages over the state-of-the-art methods. 

\subsection{Motivation}\label{model-moti}
In practice, security and efficiency are the two main obstacles to deploy secure machine learning models. 
As described in Section \ref{bg-he}, under data vertically partitioned setting, although existing HE based models have good communication efficiency since they can naturally handle high-dimensional sparse feature situation, extra information is leaked during model training procedure which causes security risk. 
Therefore, it is dangerous to deploy it in real-world applications. 
In contrast, although existing SS based methods are provable security and computaionally effecient, they cannot handle high-dimensional sparse feature situation, as is shown in Figure \ref{fig:example}, which makes them difficulty to be applied in large scale data setting. 
This is because, in practice, participants are usually large companies who have rich computation resources, and distributed computation is easily implemented inside each participant. 
However, the network bandwidth between participants are usually quite limited, which makes communication cost be the bottleneck. 
Therefore, decreasing communication cost is the key to large scale secure machine learning models. 
To combine the advantages of HE (efficiency) and SS (security), we propose \modelname, a seCure lArge-scalE SpArse logistic Regression model. 

\subsection{Secure Sparse Matrix Multiplication Protocol by Combining HE and SS}
As we described in Section \ref{pre-ml}, secure sparse matrix multiplication is the key to secure large-scale logistic regression. 

\nosection{Notations}
Before present our proposal, we first define some notations. 
Recall in Section \ref{pre-setting} that we target the setting where data are vertically partitioned by two parties. 
We use $\mathcal{A}$ and $\mathcal{B}$ to denote the two parties. 
Correspondingly, we use $\textbf{X}_a$ and $\textbf{X}_b$ to denote the features of $\mathcal{A}$ and $\mathcal{B}$, and assume $\textbf{Y}$ are the labels hold by $\mathcal{B}$. 
Let $\{pk_a, sk_a\}$ and $\{pk_b, sk_b\}$ be the HE key pairs of $\mathcal{A}$ and $\mathcal{B}$, respectively. 
Let $\llbracket x \rrbracket_a$ and $\llbracket x \rrbracket_b$ be the ciphertext of $x$ that are encrypted by using $pk_a$ and $pk_b$. 

\begin{protocol}[t]
\caption{Secure Sparse Matrix Multiplication}\label{smmhe}
\KwIn {A sparse matrix $\textbf{X}$ hold by $\mathcal{A}$, a matrix $\textbf{Y}$ hold by $\mathcal{B}$, HE key pair for $\mathcal{A}$ ($\{pk_a, sk_a\}$), HE key pair for $\mathcal{B}$ ($\{pk_b, sk_b\}$)}
\KwOut{$\textbf{Z}_1$ for $\mathcal{A}$ and $\textbf{Z}_2$ for $\mathcal{B}$ thus that $\textbf{Z}_1 + \textbf{Z}_2 = \textbf{X} \cdot \textbf{Y}$}

	$\mathcal{B}$ encrypts $\textbf{Y}$ with $pk_b$ and sends $\llbracket \textbf{Y} \rrbracket_b$ to $\mathcal{A}$ \\
	$\mathcal{A}$ calculates $\llbracket \textbf{Z} \rrbracket_b = \textbf{X} \cdot \llbracket \textbf{Y} \rrbracket_b$ \\
	$\mathcal{A}$ secretly shares $\llbracket \textbf{Z} \rrbracket_b$ using Protocol \ref{sshe}, and after that $\mathcal{A}$ gets $\textbf{Z}_1$ and $\mathcal{B}$ gets $\textbf{Z}_2$ \\
\Return $\textbf{Z}_1$ for $\mathcal{A}$ and $\textbf{Z}_2$ for $\mathcal{B}$
\end{protocol}

\begin{protocol}[t]
\caption{Secret Sharing in Homomorphically Encrypted Field}\label{sshe}
\KwIn {Homomorphically encrypted matrix $\llbracket \textbf{Z} \rrbracket_b$ for $\mathcal{A}$, HE key pair for $\mathcal{B}$ ($\{pk_b, sk_b\}$)}
\KwOut{$\left\langle\textbf{Z}\right\rangle_1$ for $\mathcal{A}$ and $\left\langle\textbf{Z}\right\rangle_2$ for $\mathcal{B}$}

	$\mathcal{A}$ locally generates share $\left\langle\textbf{Z}\right\rangle_1$ from $\mathds{Z}_\phi$ \label{sshe-step1}\\
	$\mathcal{A}$ calculates $\llbracket \left\langle\textbf{Z}\right\rangle_2 \rrbracket_b = \llbracket \textbf{Z} \rrbracket_b - \left\langle\textbf{Z}\right\rangle_1 ~\text{mod}~ \psi$ and sends $\llbracket \left\langle\textbf{Z}\right\rangle_2 \rrbracket_b$ to $\mathcal{B}$ \label{sshe-step2} \\
	$\mathcal{B}$ decrypts $\llbracket \left\langle\textbf{Z}\right\rangle_2 \rrbracket_b$ and gets $\left\langle\textbf{Z}\right\rangle_2$ \label{sshe-step3} \\
\Return $\left\langle\textbf{Z}\right\rangle_1$ for $\mathcal{A}$ and $\left\langle\textbf{Z}\right\rangle_2$ for $\mathcal{B}$
\end{protocol}

\nosection{Secure sparse matrix multiplication}
We then present a secure sparse matrix multiplication protocol in \textbf{Protocol \ref{smmhe}}. 
Given a sparse matrix $\textbf{X}$ hold by $\mathcal{A}$ and a dense matrix $\textbf{Y}$ hold by $\mathcal{B}$ 
we aim to securely calculate $\textbf{X} \cdot \textbf{Y}$ without revealing the value of $\textbf{X}$ and $\textbf{Y}$. 
In Protocol \ref{smmhe}, Line \ref{sshe-step1} shows that $\mathcal{B}$ encrypts $\textbf{Y}$ and sends it to $\mathcal{A}$. 
Line 2 is the ciphertext multiplication using additive HE, which can be significnatly speed up by parallelization as long as $\textbf{X}$ is sparse. 
Line 3 shows how to generate secrets under homomorphically encrypted field, as shown in Protocol \ref{sshe} \cite{hall2011secure}. 
Compared with the existing SS based secure matrix multiplication (Section \ref{sec-pre-ss}), the communication cost of Protocol \ref{smmhe} will be much cheaper when $\textbf{Y}$ is smaller than $\textbf{X}$, which is common in machine learning models, e.g., the model parameter vector is smaller than the feature matrix in logistic regression model. 
We present the security proof of Protocol \ref{smmhe} and Protocol \ref{sshe} in Appendix \ref{appen-b}. 




\setcounter{algocf}{0}
\begin{algorithm*}
\caption{\modelname: seCure lArge-scalE SpArse logistic Regression}\label{learning}
\KwIn {features for party $\mathcal{A}$ ($\textbf{X}_a$), features for party $\mathcal{B}$ ($\textbf{X}_b$), labels for $\mathcal{B}$ ($\textbf{y}$), HE key pair for $\mathcal{A}$ ($\{pk_a, sk_a\}$), HE key pair for $\mathcal{B}$ ($\{pk_b, sk_b\}$), max iteration number ($T$), and polynomial coefficients ($q_0, q_1, q_2$)}
\KwOut{models for party $\mathcal{A}$ ($\textbf{w}_a$) and models for party $\mathcal{B}$ ($\textbf{w}_b$)}
\underline{\textbf{Initialization:}}\\
$\mathcal{A}$ and $\mathcal{B}$ initialize their logistic regression models, i.e., $\textbf{w}_a$ and $\textbf{w}_b$, respectively \\
$\mathcal{A}$ and $\mathcal{B}$ exchange their public key $pk_a$ and $pk_b$ \\

\underline{\textbf{Secretly share models:}}\\
$\mathcal{A}$ locally generates shares $\left\langle\textbf{w}_a\right\rangle_1$ and $\left\langle\textbf{w}_a\right\rangle_2$, keeps $\left\langle\textbf{w}_a\right\rangle_1$, and sends $\left\langle\textbf{w}_a\right\rangle_2$ to $\mathcal{B}$ \label{learing-ss-a} \\
$\mathcal{B}$ locally generates shares $\left\langle\textbf{w}_b\right\rangle_1$ and $\left\langle\textbf{w}_b\right\rangle_2$, 
keeps $\left\langle\textbf{w}_b\right\rangle_2$, and sends $\left\langle\textbf{w}_b\right\rangle_1$ to $\mathcal{A}$ \label{learing-ss-b}\\

\underline{\textbf{Training model:}}\\
\For{$t=1$ to $T$}
{
	\underline{\textbf{Calculate prediction:}}\\
	$\mathcal{A}$ calculates $\left\langle\textbf{z}_a\right\rangle_1 = \textbf{X}_a \cdot \left\langle\textbf{w}_a\right\rangle_1$ \label{learning-pre-1}\\
	$\mathcal{A}$ and $\mathcal{B}$ securely calculate 
	$\left\langle\textbf{z}_a\right\rangle_2 = \textbf{X}_a \cdot \left\langle\textbf{w}_a\right\rangle_2$ 
	using Protocol \ref{smmhe}, and after that $\mathcal{A}$ gets 
	$\left\langle \left\langle\textbf{z}_a\right\rangle_2 \right\rangle_1$ 
	and $\mathcal{B}$ gets the result $\left\langle \left\langle\textbf{z}_a\right\rangle_2 \right\rangle_2$ \\

	$\mathcal{B}$ calculates $\left\langle\textbf{z}_b\right\rangle_2 = \textbf{X}_b \cdot \left\langle\textbf{w}_b\right\rangle_2$ \\

	$\mathcal{A}$ and $\mathcal{B}$ securely calculate 
	$\left\langle\textbf{z}_b\right\rangle_1 = \textbf{X}_b \cdot \left\langle\textbf{w}_b\right\rangle_1$ using Protocol \ref{smmhe}, and after that $\mathcal{A}$ gets $\left\langle \left\langle\textbf{z}_b\right\rangle_1 \right\rangle_1$ 
	and $\mathcal{B}$ gets the result $\left\langle \left\langle\textbf{z}_b\right\rangle_1 \right\rangle_2$ \label{learning-pre-4}\\
	
	$\mathcal{A}$ calculates $\left\langle\textbf{z}\right\rangle_1 = \left\langle\textbf{z}_a\right\rangle_1 + \left\langle \left\langle\textbf{z}_a\right\rangle_2 \right\rangle_1 + \left\langle \left\langle\textbf{z}_b\right\rangle_1 \right\rangle_1$, $\left\langle\textbf{z}\right\rangle_1 ^2$, and $\left\langle\textbf{z}\right\rangle_1 ^3$ and sends ciphertext $\llbracket \left\langle\textbf{z}\right\rangle_1 \rrbracket_a$, $\llbracket \left\langle\textbf{z}\right\rangle_1 ^2 \rrbracket_a$, and $\llbracket \left\langle\textbf{z}\right\rangle_1 ^3 \rrbracket_a$ to $\mathcal{B}$ \label{learning-appro-a} \\

	$\mathcal{B}$ calculates $\left\langle\textbf{z}\right\rangle_2 = \left\langle\textbf{z}_b\right\rangle_2 + \left\langle \left\langle\textbf{z}_a\right\rangle_2 \right\rangle_2 + \left\langle \left\langle\textbf{z}_b\right\rangle_1 \right\rangle_2$, $\llbracket \textbf{z} \rrbracket_a = \llbracket \left\langle\textbf{z}\right\rangle_1 \rrbracket_a + \left\langle\textbf{z}\right\rangle_2$, and $\llbracket \textbf{z}^3 \rrbracket_a = \llbracket \left\langle\textbf{z}\right\rangle_1 ^3 \rrbracket_a + 3 \llbracket \left\langle\textbf{z}\right\rangle_1 ^2 \rrbracket_a \odot \left\langle\textbf{z}\right\rangle_2 + 3 \llbracket \left\langle\textbf{z}\right\rangle_1 \rrbracket_a \odot \left\langle\textbf{z}\right\rangle_2 ^2 + \left\langle\textbf{z}\right\rangle_2 ^3$ \label{learning-appro-b} \\

	$\mathcal{B}$ calculates $\llbracket \hat{\textbf{y}} \rrbracket _a = q_0 + q_1 \llbracket \textbf{z} \rrbracket_a + q_2 \llbracket \textbf{z}^3 \rrbracket_a$, $\llbracket \textbf{e} \rrbracket _a = \llbracket \hat{\textbf{y}} \rrbracket _a - \textbf{y}$, and secretly shares $\llbracket \hat{\textbf{y}} \rrbracket_a $ using Protocol \ref{sshe}, and after that $\mathcal{A}$ gets $\left\langle \hat{\textbf{y}} \right\rangle_1$ and $\mathcal{B}$ gets $\left\langle \hat{\textbf{y}} \right\rangle_2$ \label{learning-pre-ss}

	\underline{\textbf{Calculate shared error:}}\\
	$\mathcal{A}$ calculates error $\left\langle \textbf{e} \right\rangle_1 = \left\langle \hat{\textbf{y}} \right\rangle_1$ \label{learning-err-a}\\
	$\mathcal{B}$ calculates error $\left\langle \textbf{e} \right\rangle_2 = \left\langle \hat{\textbf{y}} \right\rangle_2 - \textbf{y}$ \label{learning-err-b}\\

	\underline{\textbf{Calculate gradients:}}\\
	$\mathcal{B}$ locally calculates $\llbracket \textbf{e} \rrbracket_a ^T = \llbracket \hat{\textbf{y}} \rrbracket _a - \textbf{y}$ and $\llbracket \textbf{g}_b \rrbracket_a = \llbracket \textbf{e} \rrbracket_a ^T \cdot \textbf{X}_b$ \label{learning-grad-1}\\
	$\mathcal{B}$ secretly shares $\llbracket \textbf{g}_b \rrbracket_a$ using Protocol \ref{sshe}, and after that $\mathcal{A}$ gets $\left\langle \textbf{g}_b \right\rangle_1$ and $\mathcal{B}$ gets $\left\langle \textbf{g}_b \right\rangle_2$ \\

	$\mathcal{A}$ calculates $\left\langle\textbf{g}_a\right\rangle_1 = \left\langle \textbf{e} \right\rangle_1 ^T \cdot \textbf{X}_a$ \\
	$\mathcal{A}$ and B securely calculate $\left\langle\textbf{g}_a\right\rangle_2 = \left\langle \textbf{e} \right\rangle_2 ^T \cdot \textbf{X}_A$ using Protocol \ref{smmhe}, and after that $\mathcal{A}$ gets $\left\langle\left\langle\textbf{g}_a\right\rangle_2\right\rangle_1$ and $\mathcal{B}$ gets $\left\langle\left\langle\textbf{g}_a\right\rangle_2\right\rangle_2$ \label{learning-grad-4}\\

	\underline{\textbf{Update model:}}\\
	$\mathcal{A}$ updates $\left\langle\textbf{w}_a\right\rangle_1$ and $\left\langle\textbf{w}_b\right\rangle_1$ by $\left\langle\textbf{w}_a\right\rangle_1 \leftarrow \left\langle\textbf{w}_a\right\rangle_1 - \alpha \cdot (\left\langle\textbf{g}_a\right\rangle_1 + \left\langle\left\langle\textbf{g}_a\right\rangle_2\right\rangle_1)$ and $\left\langle\textbf{w}_b\right\rangle_1 \leftarrow \left\langle\textbf{w}_b\right\rangle_1 - \alpha \cdot \left\langle \textbf{g}_b \right\rangle_1$ \label{learning-updt-a} \\
	$\mathcal{B}$ updates $\left\langle\textbf{w}_a\right\rangle_2$ and $\left\langle\textbf{w}_b\right\rangle_2$ by $\left\langle\textbf{w}_a\right\rangle_2 \leftarrow \left\langle\textbf{w}_a\right\rangle_2 - \alpha \cdot \left\langle\left\langle\textbf{g}_a\right\rangle_2\right\rangle_2$ and $\left\langle\textbf{w}_b\right\rangle_2 \leftarrow \left\langle\textbf{w}_b\right\rangle_2 - \alpha \cdot \left\langle \textbf{g}_b \right\rangle_2$ \label{learning-updt-b}\\

}

\underline{\textbf{Reconstructing models:}}\\
$\mathcal{A}$ sends $\left\langle\textbf{w}_b\right\rangle_1$ to $\mathcal{B}$ \label{learning-rec-a} \\
$\mathcal{B}$ sends $\left\langle\textbf{w}_a\right\rangle_2$ to $\mathcal{A}$ \label{learning-rec-b} \\
$\mathcal{A}$ reconstructs $\textbf{w}_a = \left\langle\textbf{w}_a\right\rangle_1 + \left\langle\textbf{w}_a\right\rangle_2$ \label{learning-rec-c} \\
$\mathcal{B}$ reconstructs $\textbf{w}_b = \left\langle\textbf{w}_b\right\rangle_1 + \left\langle\textbf{w}_b\right\rangle_2$ \label{learning-rec-d} \\

\Return models for party $\mathcal{A}$ ($\textbf{w}_a$) and models for party $\mathcal{B}$ ($\textbf{w}_b$)
\end{algorithm*}

\subsection{Secure Large-Scale Logistic Regression}\label{sec-model-lr}
With secure sparse matrix multiplication protocol in hand, we now present \modelname, a seCure lArge-scalE SpArse logistic Regression model, in Algorithm \ref{learning}, where we omit mod (refers to Section \ref{sec-pre-ss}), mini-batch (refers to Section \ref{pre-ml}), and regularizations for conciseness. 
The basic idea is that $\mathcal{A}$ and $\mathcal{B}$ secretly \textbf{share} the models between two parties (Line \ref{learing-ss-a} and Line \ref{learing-ss-b}), so that models are always secret shares during model training, and finally \textbf{reconstruct} models after training (Line \ref{learning-rec-a}-Line \ref{learning-rec-d}). 
Meanwhile, $\mathcal{A}$ and $\mathcal{B}$ keep their private features and labels.  
During model training, we calculate the shares of $\textbf{X} \cdot \textbf{w}$ using Protocol \ref{smmhe} (Line \ref{learning-pre-1}-\ref{learning-pre-4}), and approximate the logistic function using Minimax approximation (Line \ref{learning-appro-a} and Line \ref{learning-appro-b}), as described in Section \ref{pre-appro}. 
After it, $\mathcal{B}$ calculates the prediction in ciphertext and secretly shares it using Protocol \ref{sshe} (Line \ref{learning-pre-ss}). 
$\mathcal{A}$ and $\mathcal{B}$ then calculate the shares of error (Line \ref{learning-err-a} and Line \ref{learning-err-b}) and calculate the shares of the gradients using Protocol \ref{smmhe} and Protocol \ref{sshe} (Line \ref{learning-grad-1}-\ref{learning-grad-4}). 
They finally update the shares of their models using gradient descent (Line \ref{learning-updt-a}-\ref{learning-updt-b}). 
During model training, private features and labels are kept by participants themselves, while models and gradients are either \textit{secretly shared} or \textit{homomorphically encrypted}. 
Algorithm \ref{learning} is exactly a classic secure two-party computation if one takes it as a function. 

\subsection{Advantages of CAESAR}\label{sec-model-adv}
We now analyze the advantages of \modelname~over existing secret sharing based secure logistic regression protocols \cite{mohassel2017secureml,demmler2015aby}. 
It can be seen from Algorithm \ref{learning} that, for \modelname, in each mini-batch (iteration), the communication complexity between $\mathcal{A}$ and $\mathcal{B}$ is $O(7|\textbf{B}|+2d)$. That is, $O(7n+2nd/|\textbf{B}|)$ for passing the dataset once (one epoch), where $|\textbf{B}|$ is batch size, $n$ is sample size, and $d$ is the feature size of both parties. 
In comparison, for the existing secret sharing based secure protocols, e.g., SecureML \cite{mohassel2017secureml} and ABY \cite{demmler2015aby}, the communication complexity between $\mathcal{A}$ and $\mathcal{B}$ is $O(4nd)$ for each epoch during the online phase, ignoring the time-consuming offline circuit and Beaver's triplet generation phase \cite{beaver1991efficient}. 
Our protocol has much less communication costs than existing secure LR protocols, especially when it refers to large scale datasets. 

\section{Implementation}\label{sec-impl}
In this section, we first describe the motivation of our implementation, and then present the detailed implementation of \modelname. 

\begin{figure}[t]
\centering 
\includegraphics[width=4.5cm]{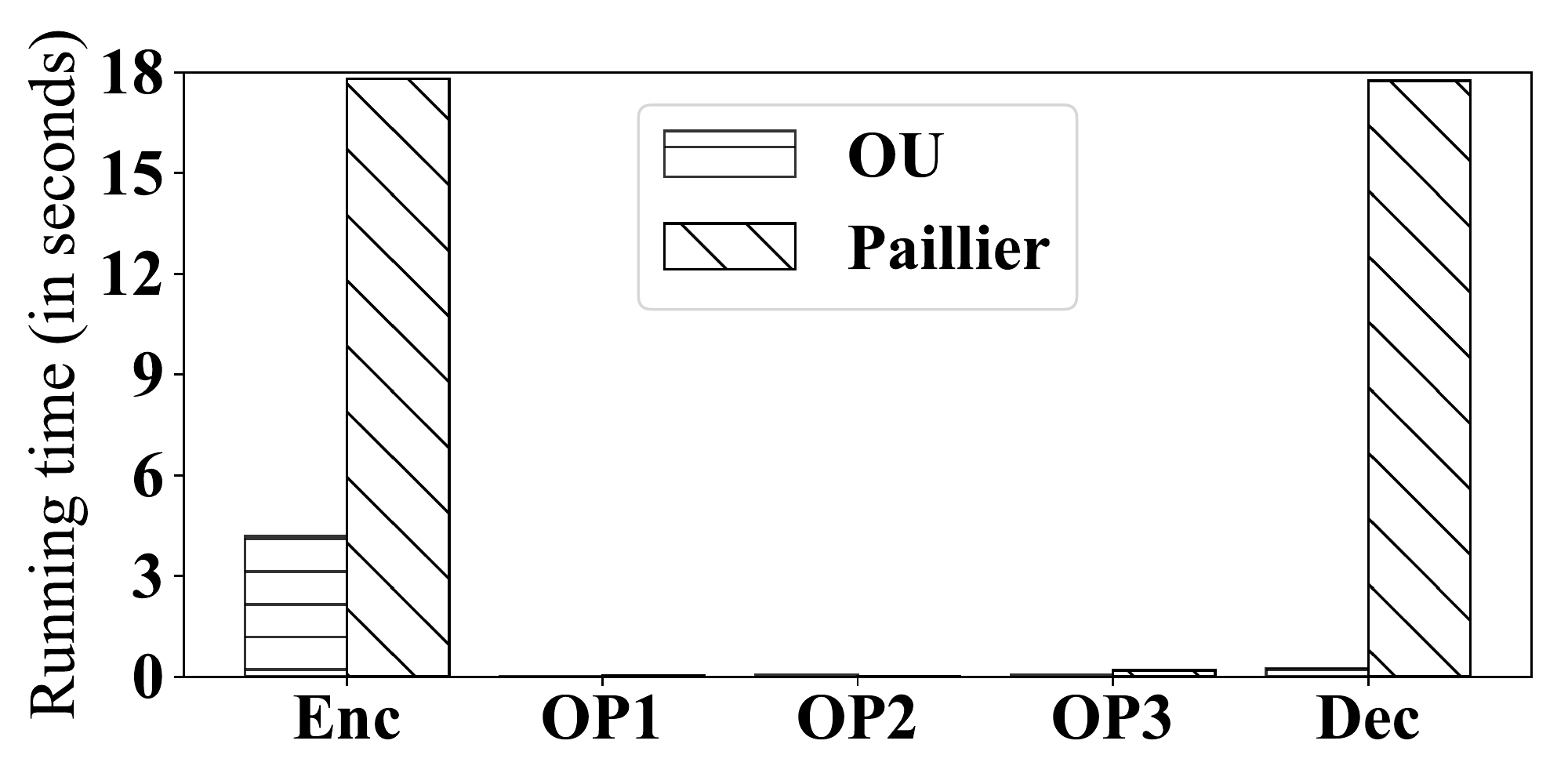} 
\vskip -0.17in
\caption{Running time comparison of different computation types for OU and Paillier. }
\label{fig:he-compare}
\vskip -0.07in
\end{figure}

\begin{figure}[t]
\centering 
\includegraphics[width=\columnwidth]{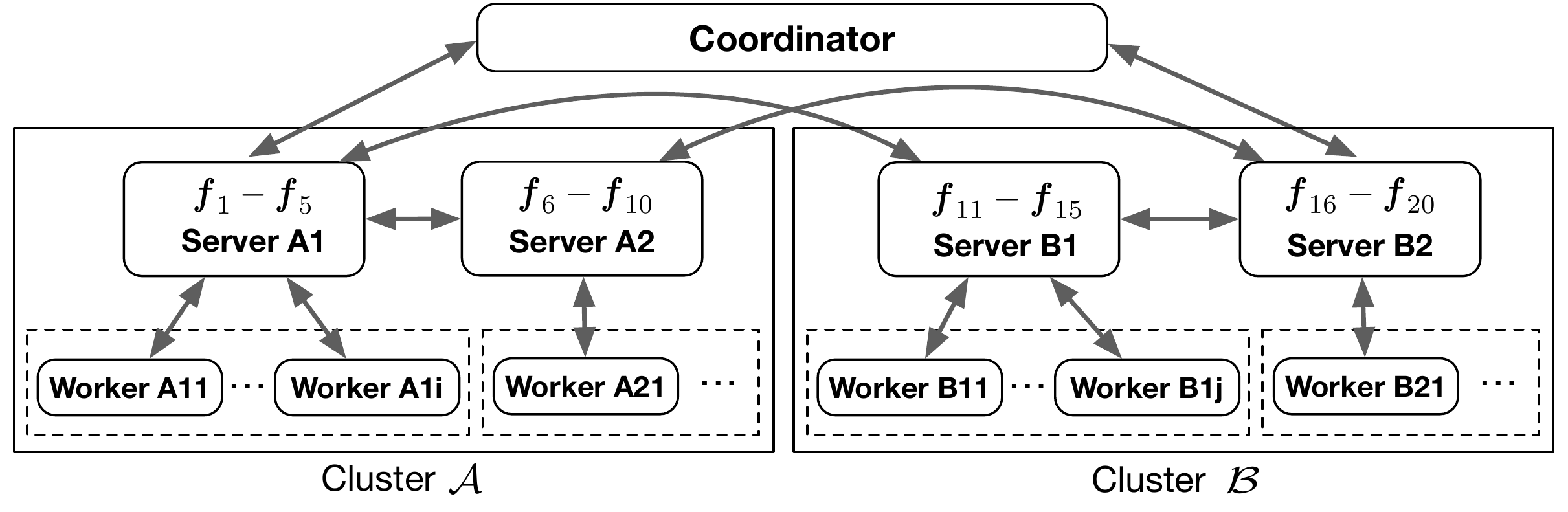} 
\vskip -0.15in
\caption{Implementation framework of \modelname. }
\label{fig:imple}
\vskip -0.1in
\end{figure}

\subsection{Motivation}

Our proposed \modelname~has overwhelming advantages against existing secret sharing based protocols in terms of communication efficiency, as we have analyzed in Section \ref{sec-model-lr}. Although \modelname~ involves additional HE operations, this can be solved by distributed computations. 
To help design reasonable distributed computation framework, we first analyze the running time of different computations of additive HE. 
We choose two additive HE methods, i.e., OU \cite{okamoto1998new} and Paillier \cite{paillier1999public}. They have five types of computations, i.e., encryption (\textbf{Enc}), decryption (\textbf{Dec}), and three types of homomorphic operations: \textbf{OP1: }$\llbracket x+y \rrbracket =x+\llbracket y \rrbracket$, \textbf{OP2: }$\llbracket x+y \rrbracket=\llbracket x \rrbracket+\llbracket y \rrbracket$, and \textbf{OP3: }$\llbracket x \cdot y \rrbracket = x \cdot \llbracket y \rrbracket$. 
We run these computations 1,000 times and report their running time in Figure \ref{fig:he-compare}. 
From it, we can find that OU has better performance than Paillier, and \textbf{Enc} is the most time-consuming computation type. 
Therefore, improving the encryption efficiency is the key to distributed implementation. 

\subsection{Distributed Implementation}

\nosection{Overview}
Overall, our designed implementation framework of \modelname~is comprised of a coordinator and two clusters on both participants' side, as shown in Figure \ref{fig:imple}. 
The coordinator controls the start and terminal of the clusters based on a certain condition, e.g., the number of iterations. Each cluster is also a distributed learning system, which consists of \textbf{servers} and \textbf{workers} and is maintained by participant ($\mathcal{A}$ or $\mathcal{B}$) itself. 

\nosection{Vertically distribute data and model on servers}
To support high-dimensional features and the corresponding model parameters, we borrow the idea of distributed data and model from parameter server \cite{li2014scaling,zhou2017kunpeng}. 
Specifically, each cluster has a group of \textbf{servers} who split features and models vertically, as shown in Figure \ref{fig:imple}, where each server has 5 features for example. The label could be held by either $\mathcal{A}$ or $\mathcal{B}$, and we omit it in Figure \ref{fig:imple} for conciseness. 
Note that $\mathcal{A}$ and $\mathcal{B}$ should have the same number of servers, so that each server pair, e.g., Server \textbf{A1} and Server \textbf{B1}, learn the model parameters using Algorithm \ref{learning}. 
In each mini-batch, all the server pairs of $\mathcal{A}$ and $\mathcal{B}$ first calculate partial predictions in parallel, and then, servers in each cluster communicate with each other to get the whole predictions. 

\nosection{Distribute encryption operations on workers}
During model training for the server pairs, when it involves encryption operation, each server distributes its plaintext data to \textit{workers} for distributed encryption. 
After workers finish encryption, they send the ciphertext data to the corresponding server for successive computations. Take Line \ref{learning-appro-a} in Algorithm \ref{learning} for example, each server of $\mathcal{A}$ (e.g., Server \textbf{A1}) sends partial plaintext data, i.e., $\left\langle\textbf{z}\right\rangle_1 $, $\left\langle\textbf{z}\right\rangle_1 ^2$, and $\left\langle\textbf{z}\right\rangle_1 ^3$, to its workers (Worker \textbf{A11} to Worker\textbf{A1i}) for encryption, and then these workers send $\llbracket \left\langle\textbf{z}\right\rangle_1 \rrbracket_a$, $\llbracket \left\langle\textbf{z}\right\rangle_1 ^2 \rrbracket_a$, and $\llbracket \left\langle\textbf{z}\right\rangle_1 ^3 \rrbracket_a$ back to Server \textbf{A1}. The communication cost in each cluster is cheap, since they are in a local area network. 
Moreover, two clusters communicate shared or encrypted information with each other to finish model learning, following Algorithm \ref{learning}. 

By vertically distributing data and model on servers, and distributing encryption operations on workers, our implementation can scale to large datasets.


\begin{figure*}[h]
\centering
\subfigure[\emph{Bandwidth=10Mbps}]{ \includegraphics[width=4.3cm]{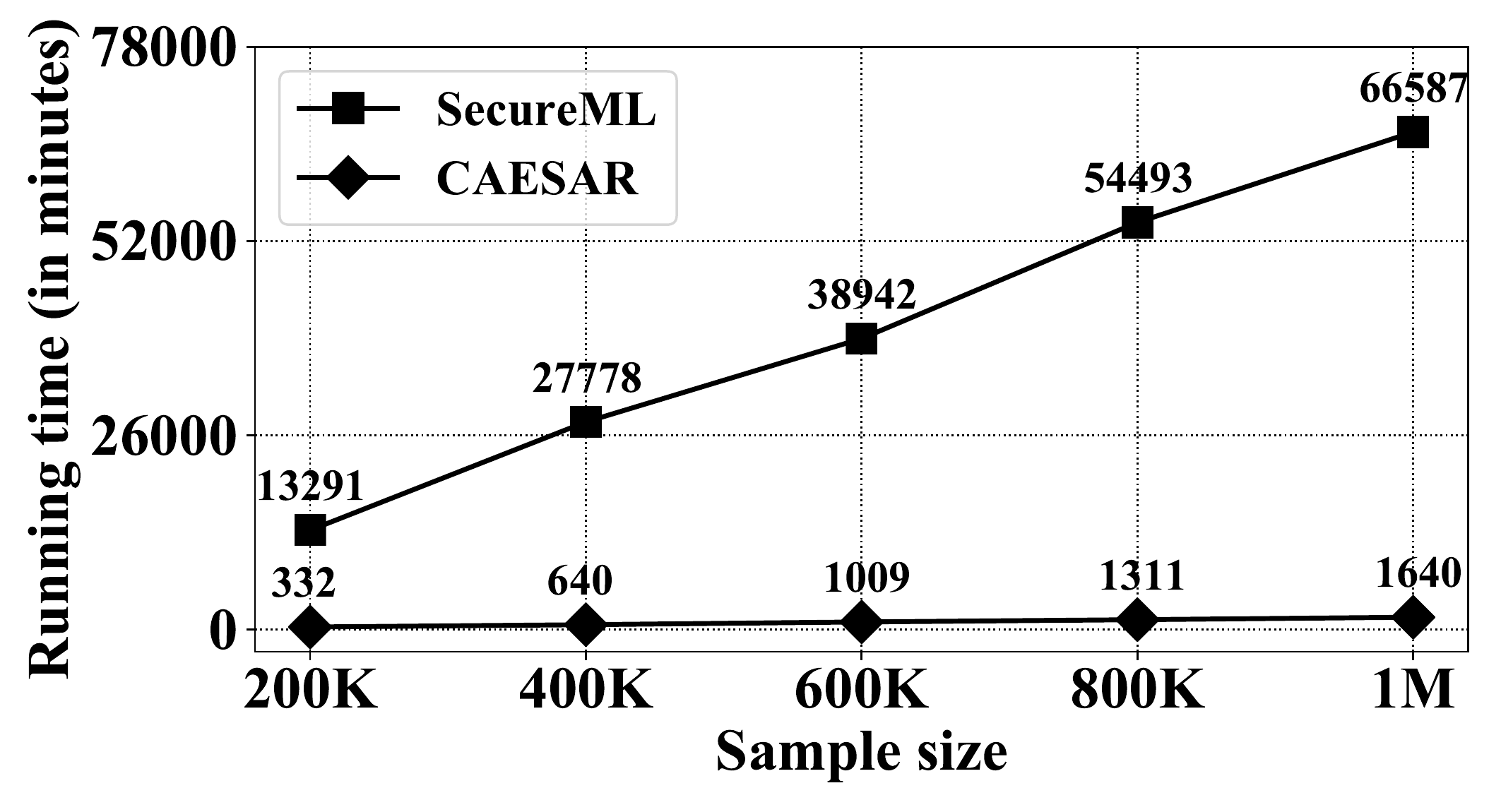}}
\subfigure[\emph{Bandwidth=20Mbps}] { \includegraphics[width=4.3cm]{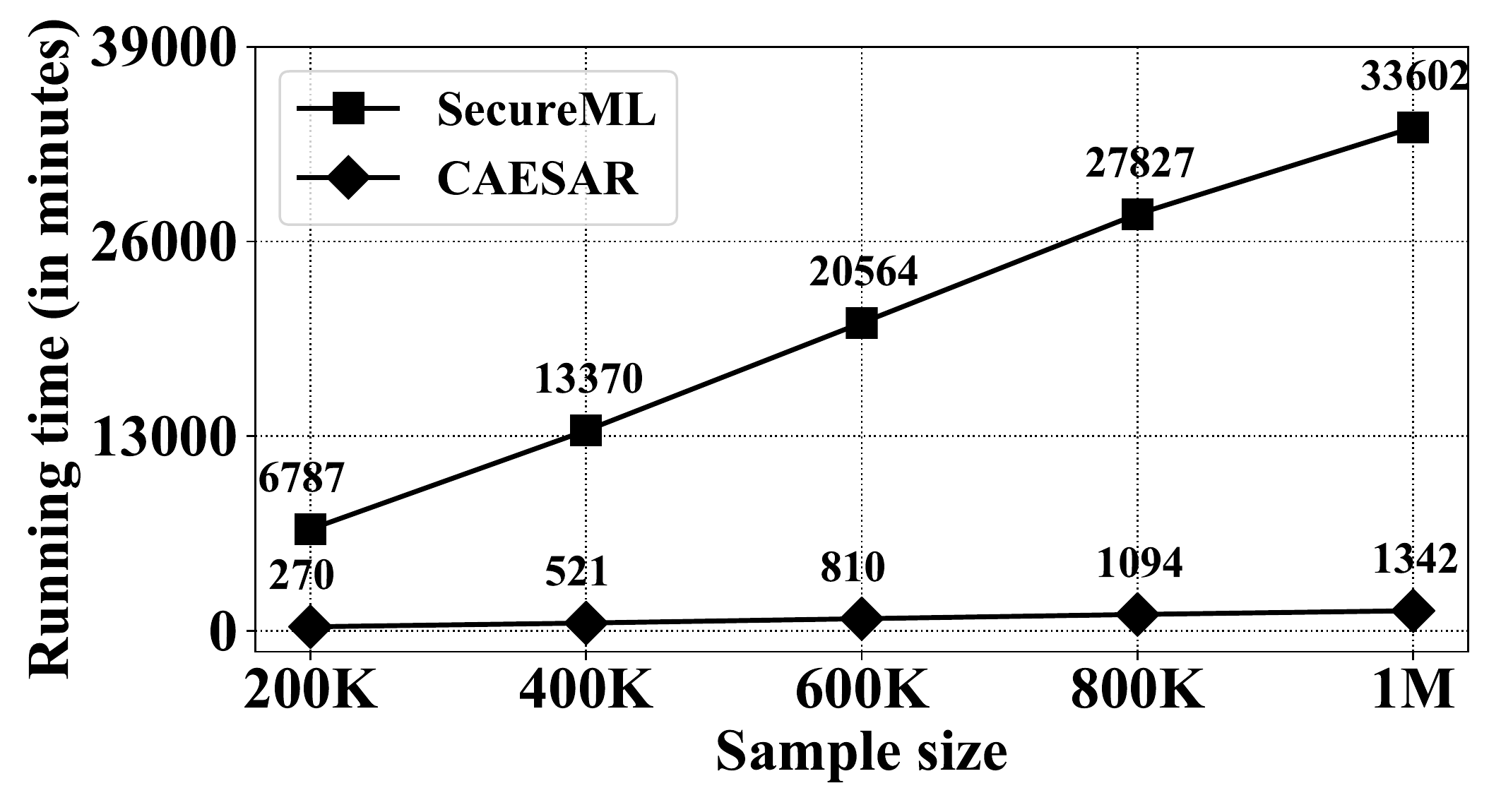}}
\subfigure[\emph{Bandwidth=30Mbps}] { \includegraphics[width=4.3cm]{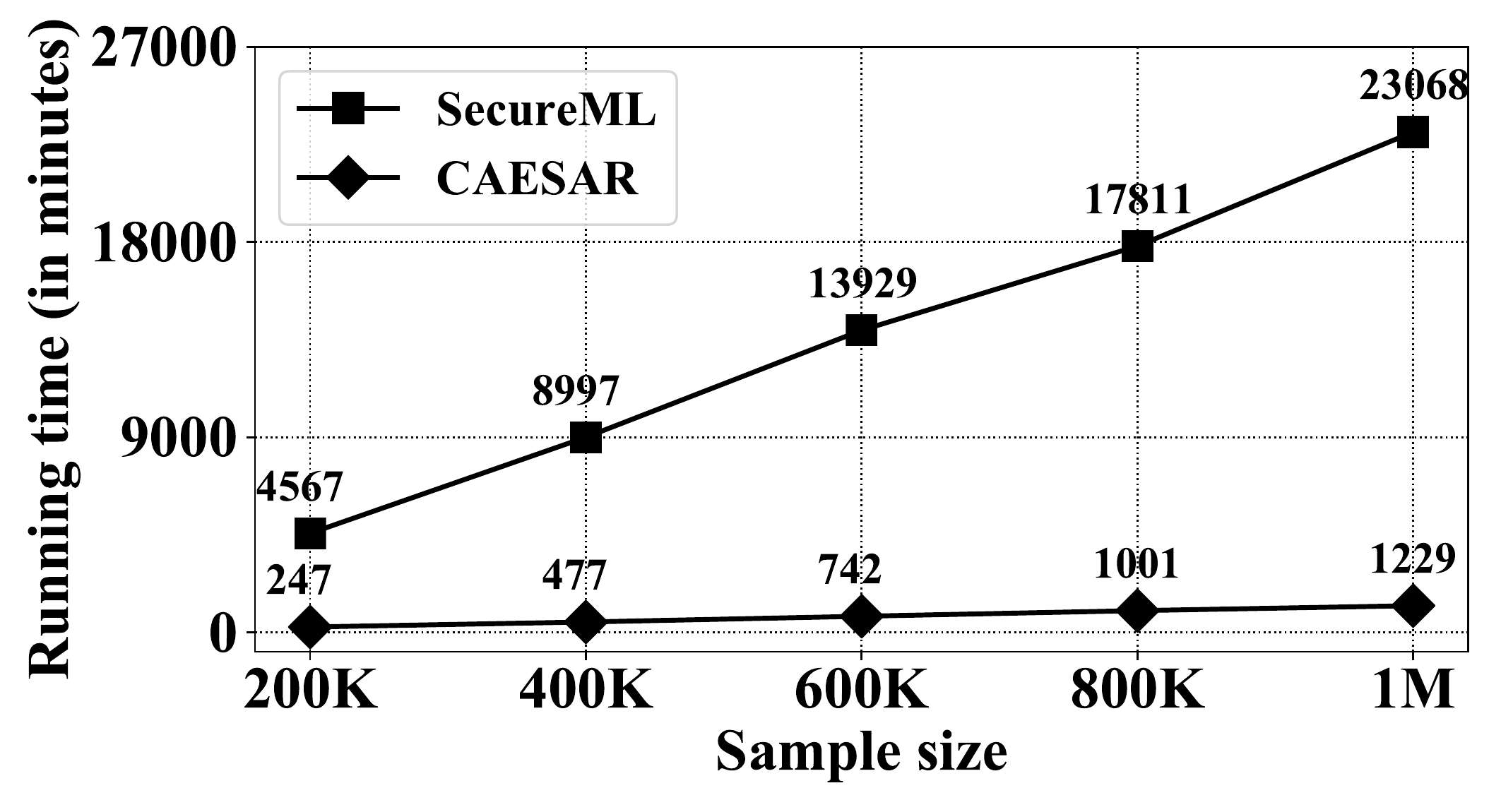}}
\subfigure[\emph{Bandwidth=40Mbps}] { \includegraphics[width=4.3cm]{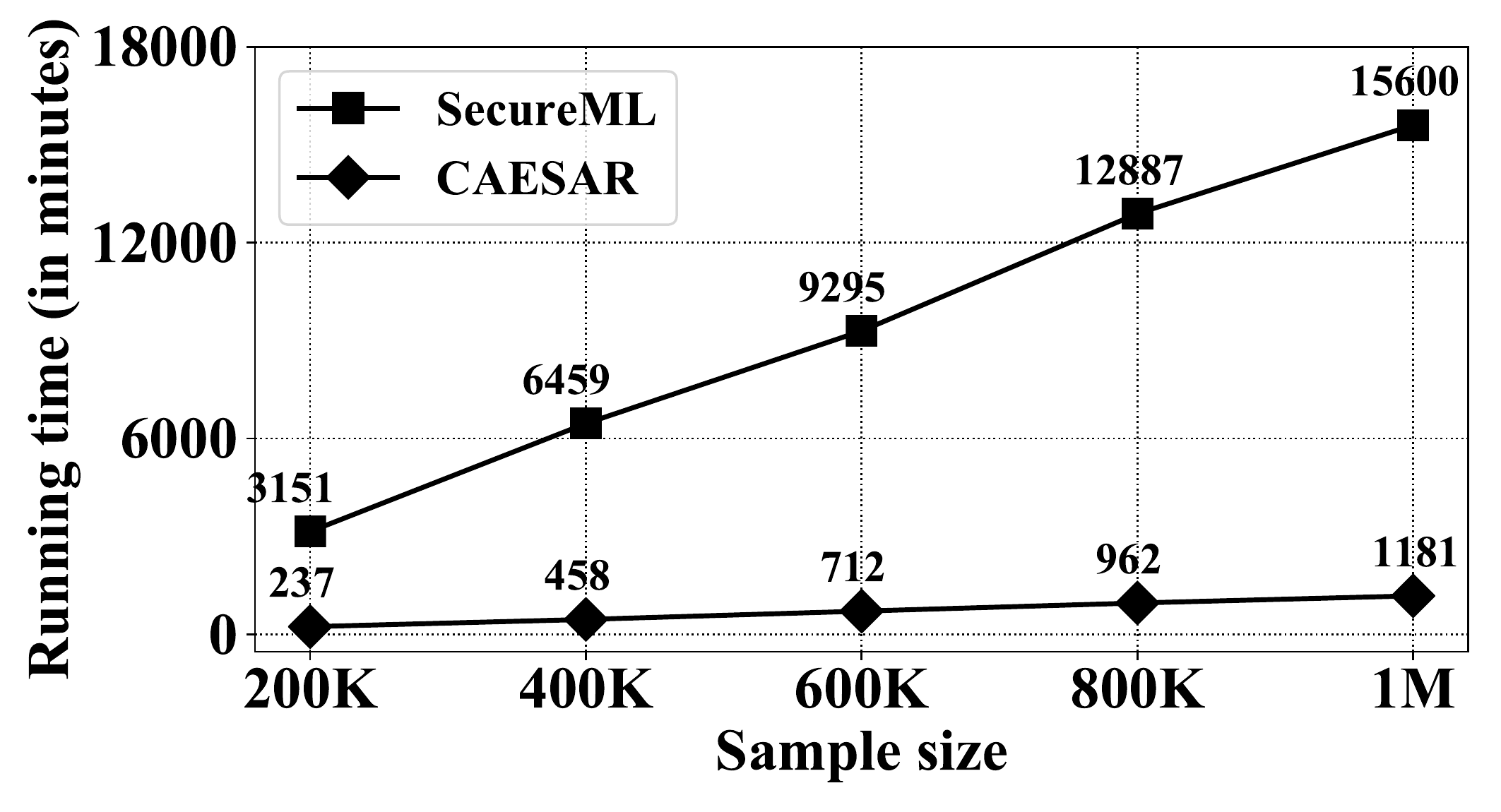}}
\vskip -0.15in
\caption{Running time (per epoch) comparison with respect to sample size by varying bandwidth (batch size = 1,024).}
\label{fig-comp1}
\vskip -0.1in
\end{figure*}

\begin{figure*}[h]
\centering
\subfigure[\emph{Bandwidth=10Mbps}]{ \includegraphics[width=4.3cm]{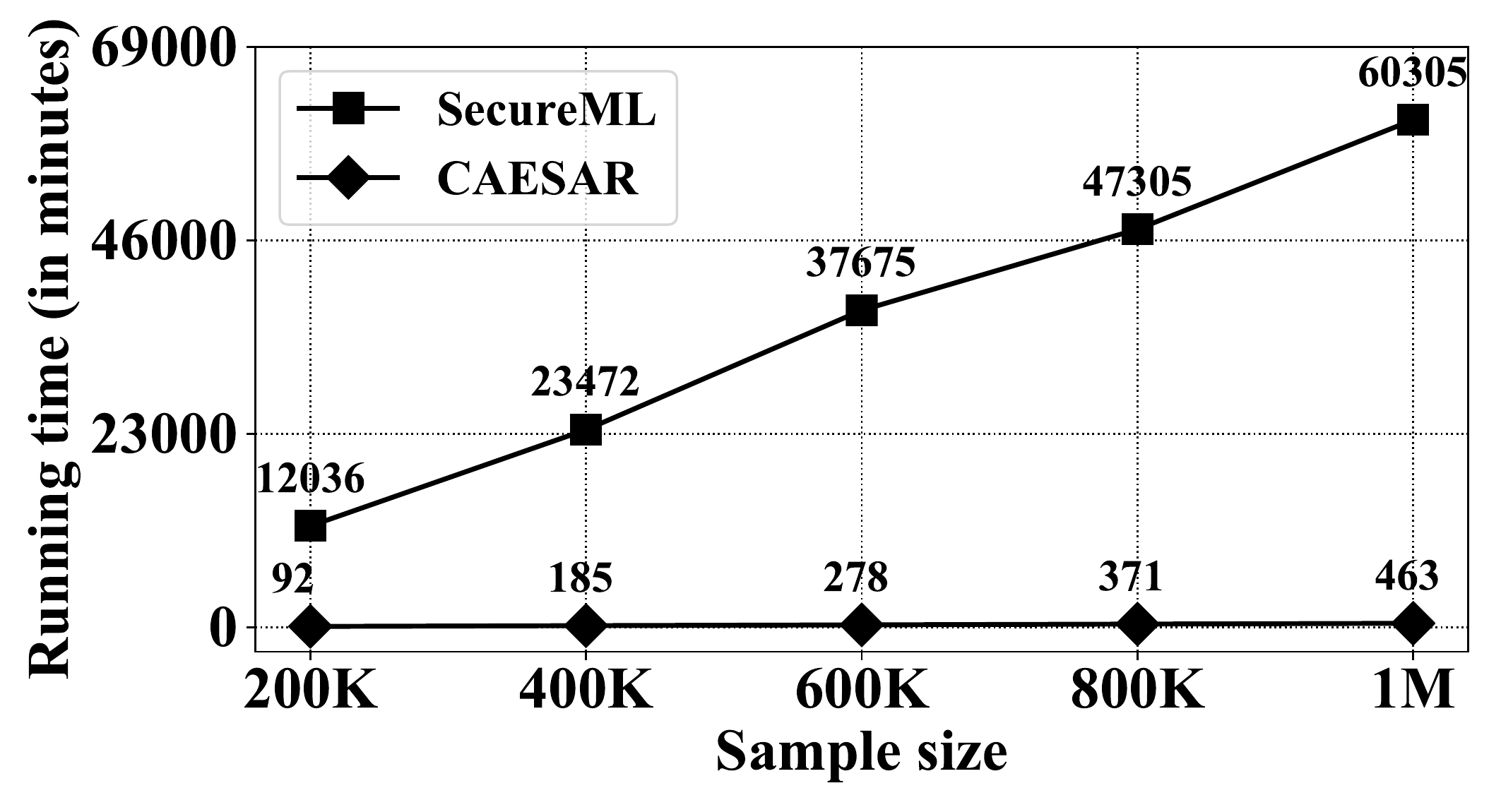}}
\subfigure[\emph{Bandwidth=20Mbps}] { \includegraphics[width=4.3cm]{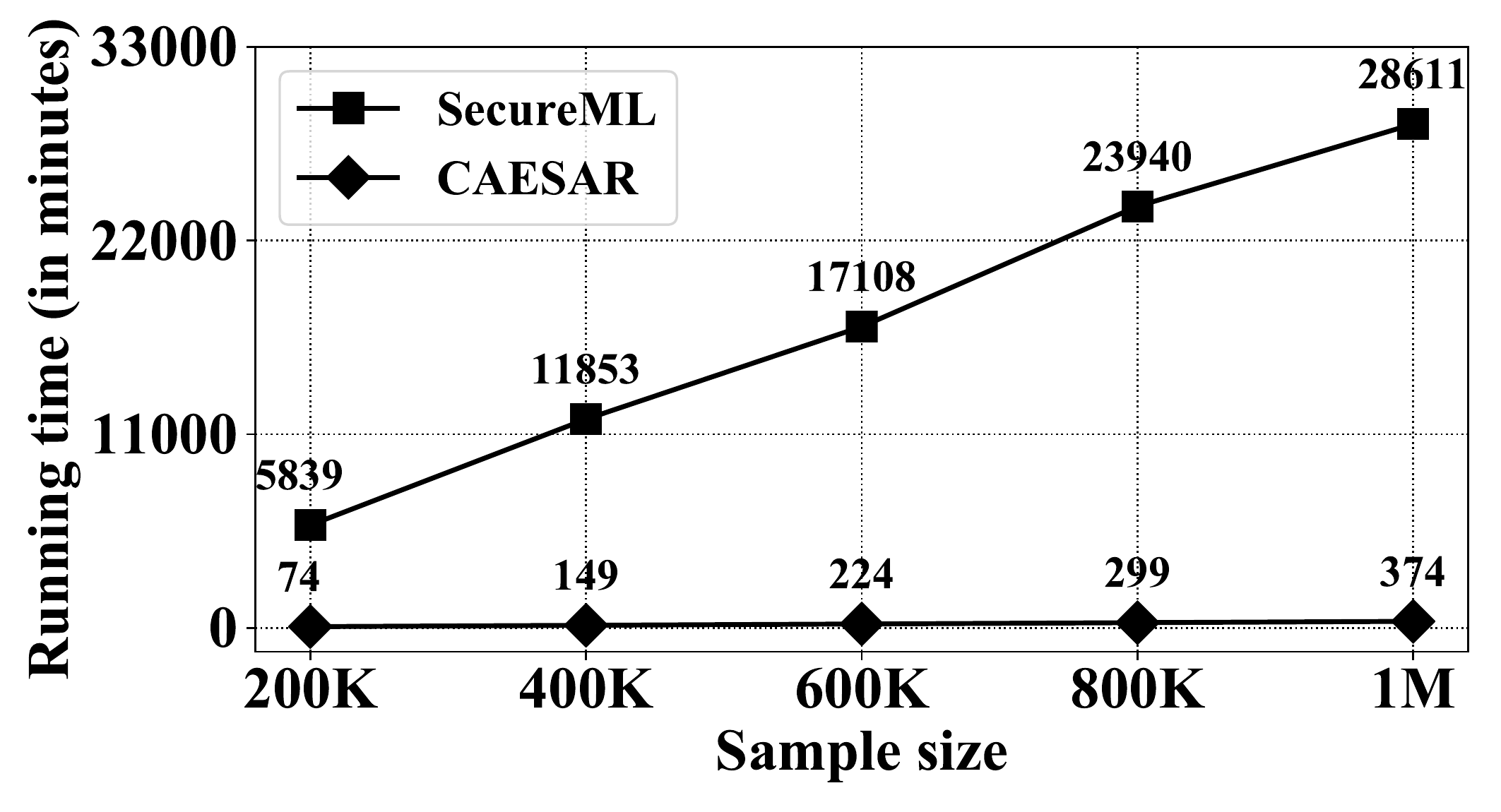}}
\subfigure[\emph{Bandwidth=30Mbps}] { \includegraphics[width=4.3cm]{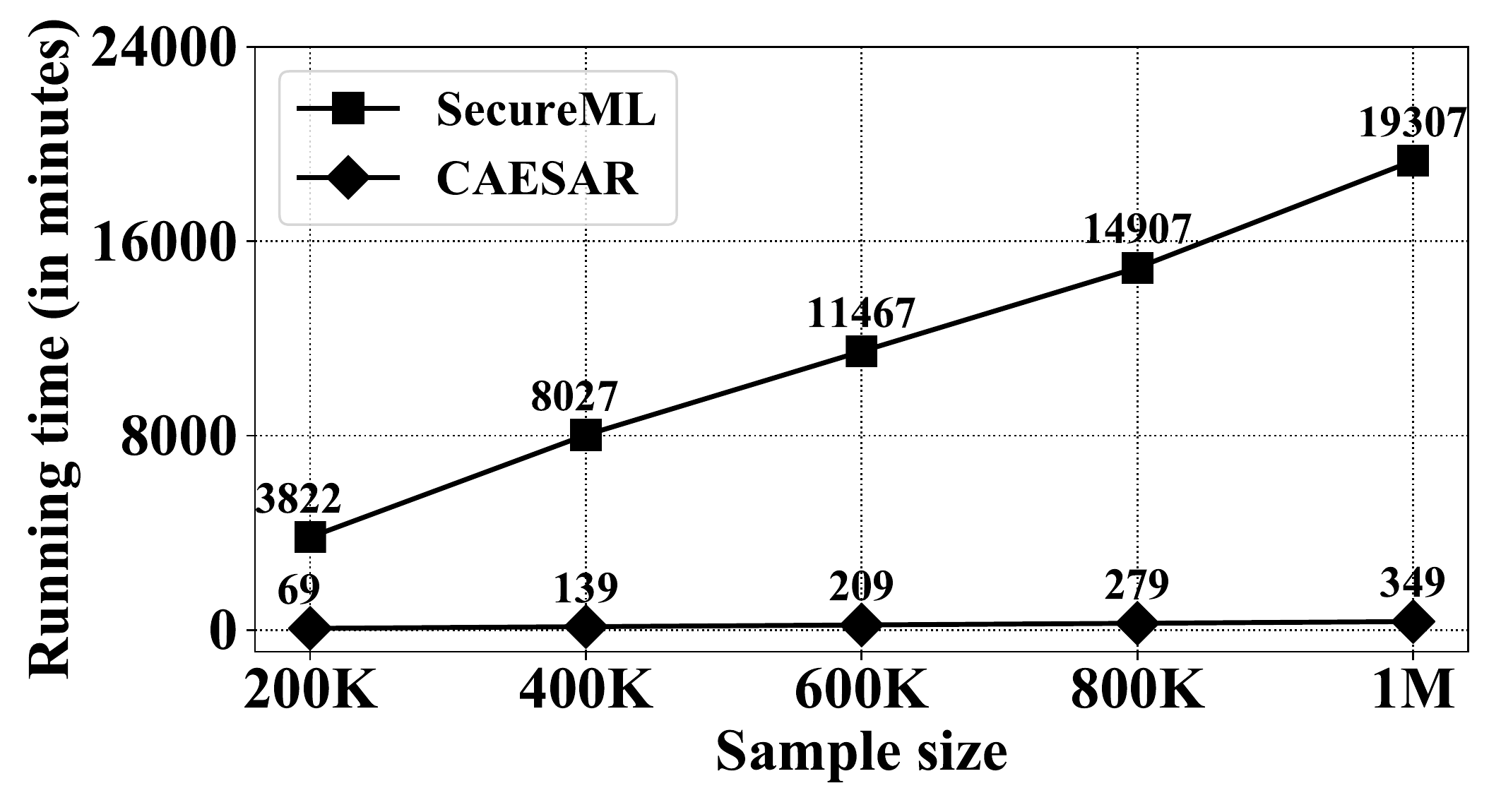}}
\subfigure[\emph{Bandwidth=40Mbps}] { \includegraphics[width=4.3cm]{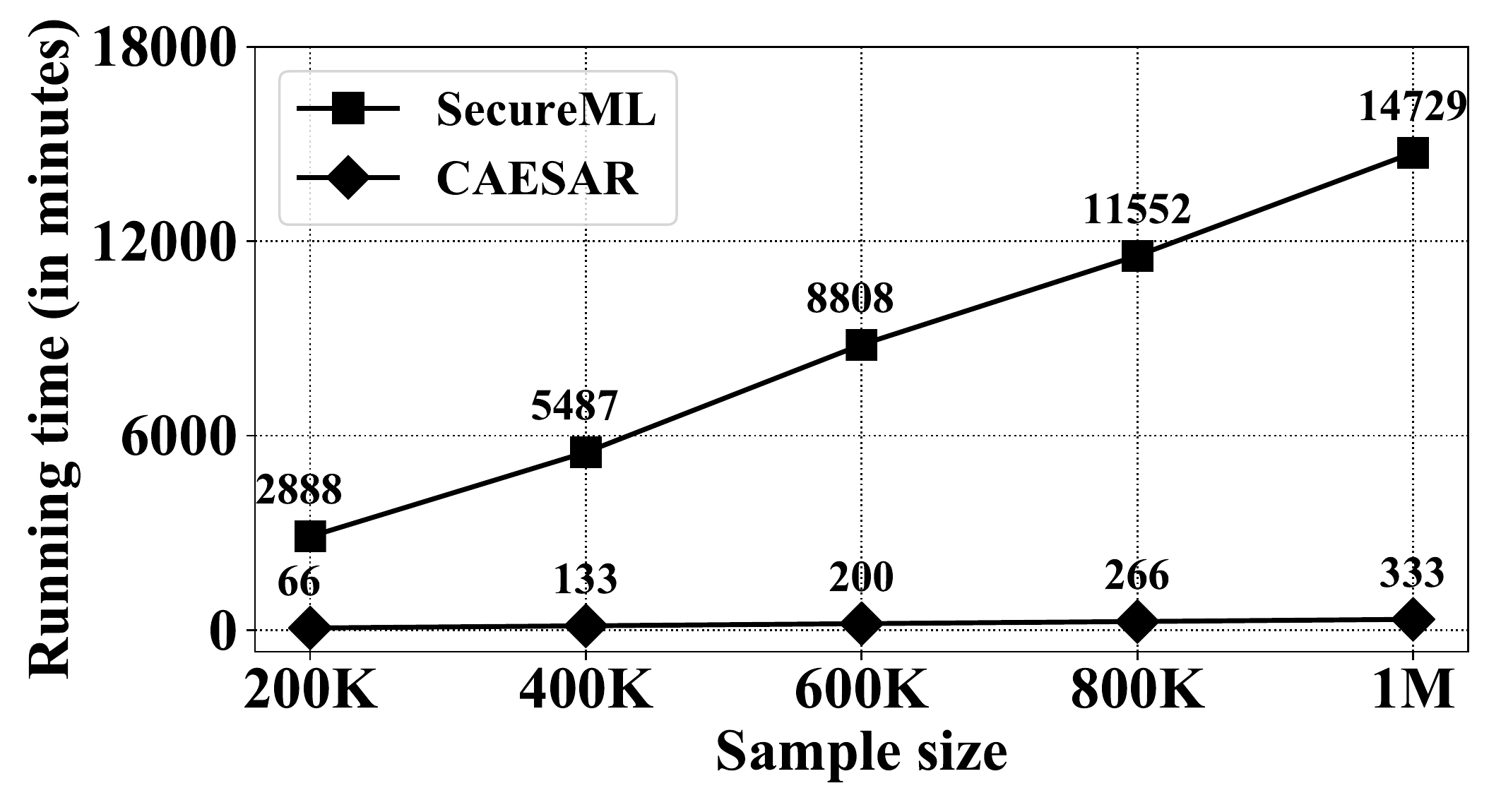}}
\vskip -0.15in
\caption{Running time (per epoch) comparison with respect to sample size by varying bandwidth (batch size = 4,096).}
\label{fig-comp2}
\vskip -0.1in
\end{figure*}

\section{Deployment and Applications}\label{sec-exp}	
In this section, we deploy \modelname~into a risk control task, and conduct comprehensive expreiments on it to study the effectiveness of \modelname.

\subsection{Experiment Setup}

\nosection{Scenario} 
\companyname~provides online payment services, whose customers include both individual users and large merchants. 
Users can make online transactions to merchants through \companyname, and to protect user's property, controlling transaction risk is rather important to both \companyname~ and its large merchants. Under such scenario, rich features, including user feature in \companyname, transaction (context) features in \companyname, and user feature in merchant, are the key to build intelligent risk control models. 
However, due to the data isolation problem, these data cannot be shared with each other directly. 
To build a more intelligent risk control system, we deploy \modelname~for \companyname~ and the merchant to collaborately build a secure logistic regression model due to the requirements of robustness and interpretability. 

\nosection{Datasets} 
We use the real-world dataset in the above scenario, where \companyname~ has 30,100 features and label and the merchant has 70,100 features, and the feature sparsity degree is about 0.02\%. 
Among them, \companyname~mainly has transaction features (e.g., transaction amount) and partial user feature (e.g., user age), while the merchant 
has the other partial user behavior features (e.g., visiting count). 
The sparse features mainly come from the incomplete user profiles or feature engineering such as one-hot. 
The whole dataset has 1,236,681 samples and among which there are 1,208,569 positive (normal) samples and 28,112 negative (risky) samples. We split the dataset into two parts based on the timeline, the 80\% earlier happened transactions are taken as training dataset while the later 20\% ones are taken as test dataset. 

\nosection{Metrics} 
We adopt four metrics to evaluate model performance for risk control task, i.e., 
(1) Area Under the ROC Curve (AUC), (2) Kolmogorov-Smirnov (KS) statistic, (3) F1 value, and (4) recall of the 90\% precision (Recall@0.9precision), i.e., the recall value when the classification model reaches 90\% precision. 
These four metrics evaluate a classification model from different aspects, the first three metrics are commonly used in literature, and the last metric is commonly used in industry under imbalanced tasks such as risk control and fraud detection. 
For all the metrics, the bigger values indicate better model performance. 

\nosection{Comparison methods} 
To test the \textit{effectiveness} of \modelname, we compare it with the plaintext logistic regression model using \companyname's features only, and the existing SS based secure logistic regression, i.e., SecureML \cite{mohassel2017secureml}. 
To test the \textit{efficiency} of \modelname, we compare it with SecureML \cite{mohassel2017secureml}. SecureML is based on secret sharing, and thus cannot handle high-dimensional sparse features, as we have described in Figure \ref{fig:example}. 
Note that, we cannot empirically compare the performance of \modelname~with the plaintext logistic regression using mixed plaintext data, since the data are isolated. 

\nosection{Hyper-parameters} 
We choose OU as the HE method, and set key length to 2,048. 
We set $l=64$ and set $\psi$ to be a large number with longer than 1,365 bits (2/3 of the key length). 
We fix the server number to 1 since it is suitable for our dataset, and vary the number of workers and bandwidth to study their effects on \modelname. 

\begin{figure*}
\centering
\subfigure [\emph{Effect of worker number}]{ \includegraphics[width=4.3cm]{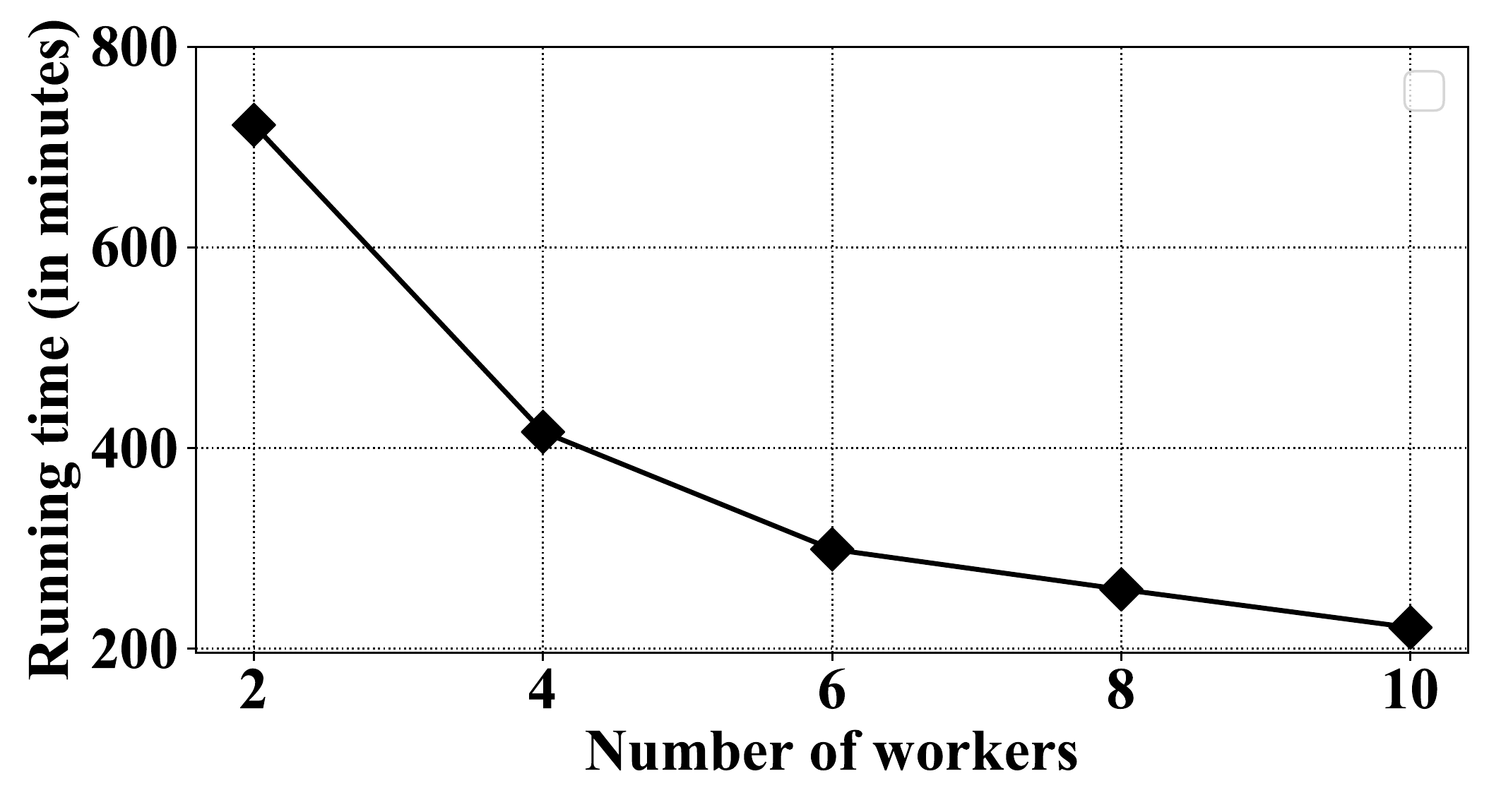}}
\subfigure[\emph{Effect of feature number}] { \includegraphics[width=4.3cm]{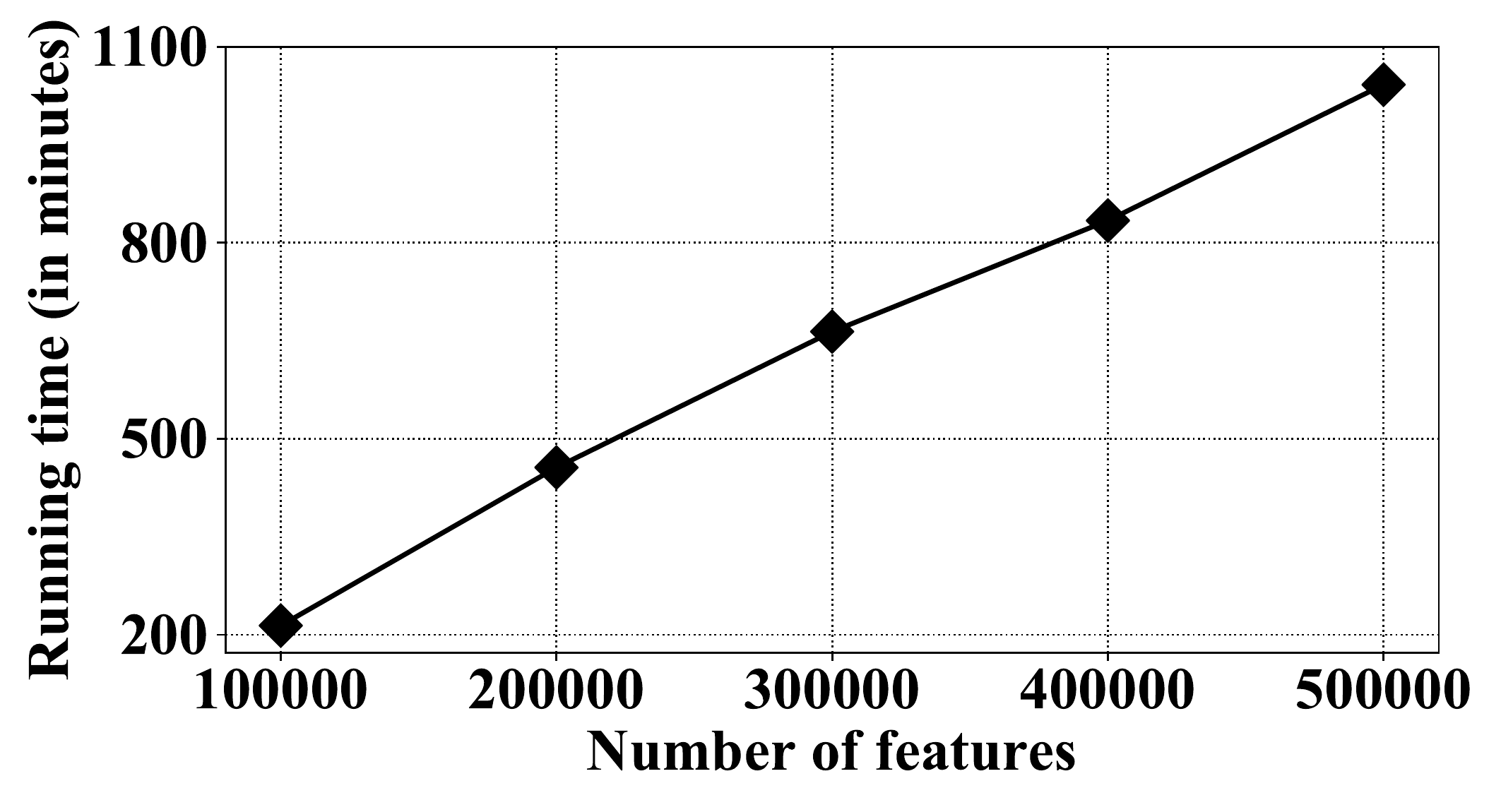}}
\subfigure[\emph{Effect of batch size}] { \includegraphics[width=4.3cm]{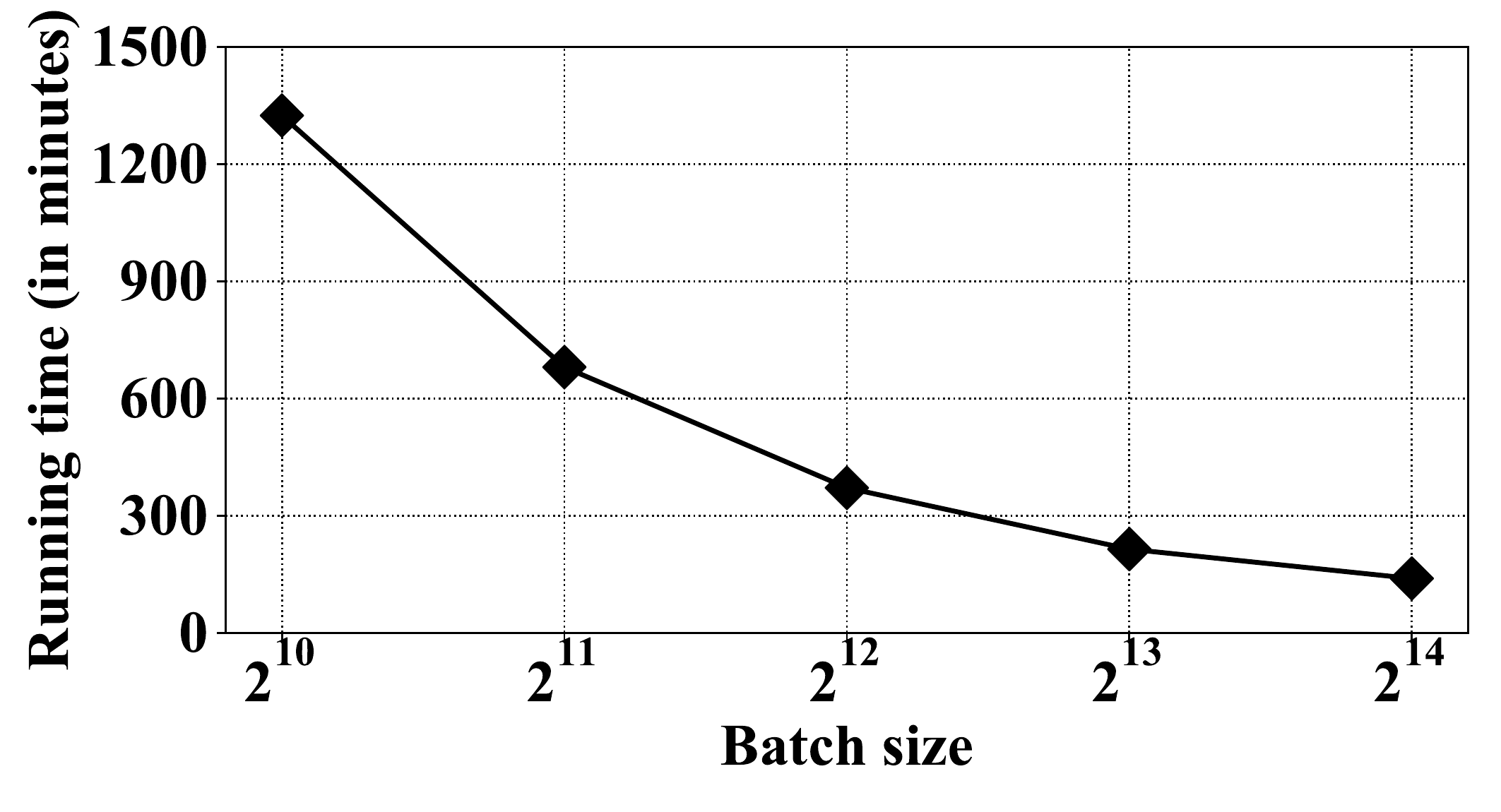}}
\subfigure[\emph{Effect of network bandwidth}] { \includegraphics[width=4.3cm]{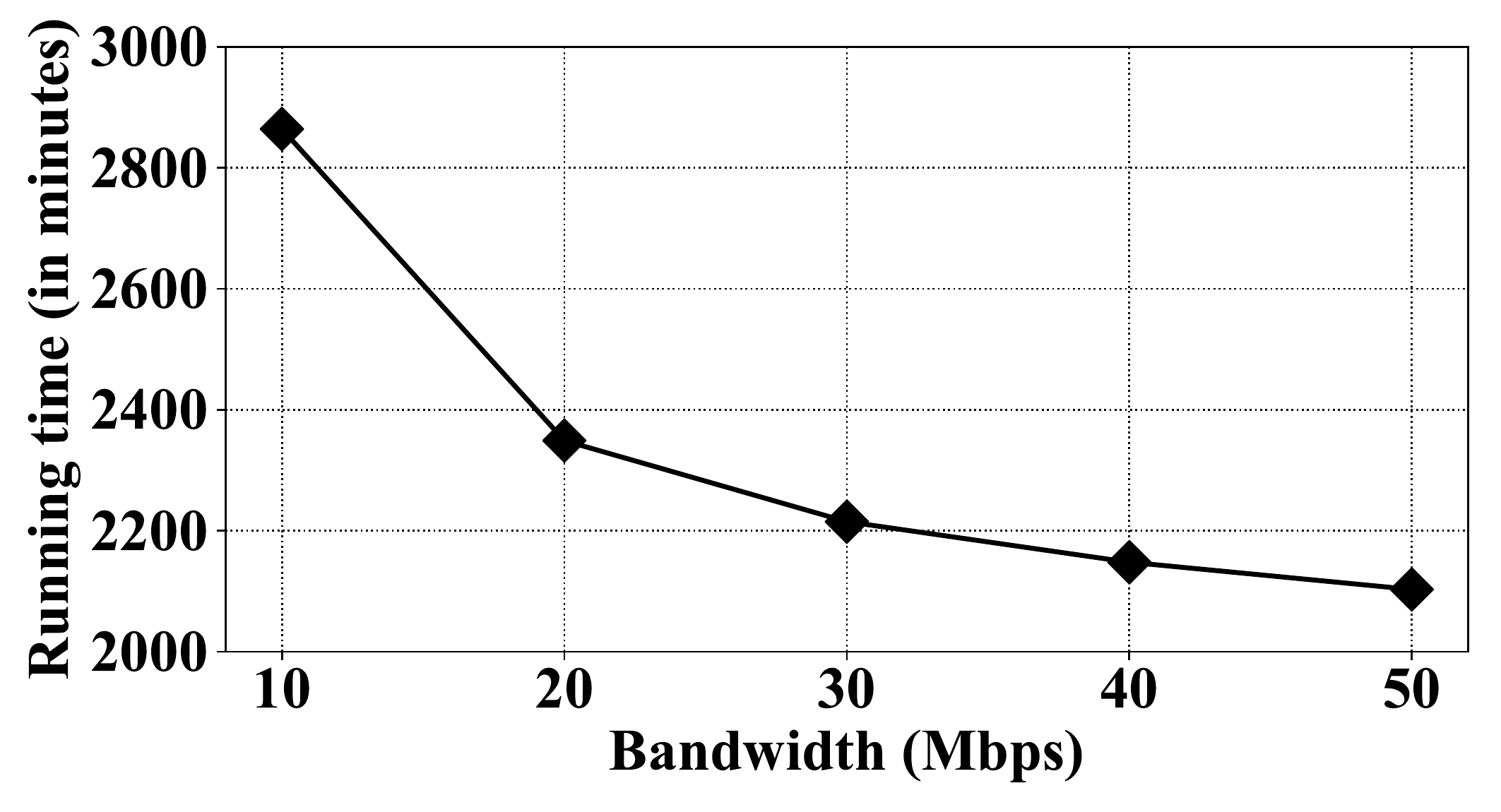}}
\vskip -0.15in
\caption{Effects of parameters on \modelname.}
\label{fig-param-effect}
\vskip -0.1in
\end{figure*}

\begin{table}
\centering\small
\caption{Comparison results }
\vskip -0.15in
\label{table-compare}
\begin{tabular}{ccccc}
  \toprule
  Metric & AUC & KS & F1 & Recall@0.9precision \\
  \midrule
  \companyname-LR & 0.9862 & 0.9018 & 0.5350 & 0.2635 \\
  SecureML & 0.9914 & 0.9415 & 0.6167 & 0.3598 \\
  \modelname & \textbf{0.9914} & \textbf{0.9415} & \textbf{0.6167} & \textbf{0.3598} \\
  \bottomrule
\end{tabular}
\end{table}

\subsection{Comparison Results}

\nosection{Effectiveness} 
We first compare \modelname~with plaintext logistic regression using \companyname's data only (\companyname-LR) to test its effectiveness. 
Traditionally, \companyname~ can build logistic regression model using its plaintext features only. 
With secure logistic regression models, i.e., SecureML and \modelname, \companyname~ can build better logistic regression model together with the merchant, without compromising their private data. 
We summarize their comparison results in Table \ref{table-compare}. 
From it, we can observe that \companyname-LR can already achieve satisfying performance. 
However, we also find that SecureML and \modelname~have the same performance, i.e., they consistently achieve much better performance than \companyname-LR. 
Take Recall@0.9precision---the most practical metric in industry---for example, \modelname~increases the recall rate of Ant-LR as high as 36.55\% while remaining the same precision (90\%). 
This means that \modelname~captures 36.55\% more risky transactions than the traditional \companyname-LR model while keeping the same accuracy, which is quite a high improvement in risk control task. 
The result is easy to explain: more valuable features will naturally improve risk control ability, which indicates the effectiveness of \modelname. 

\nosection{Efficiency} 
We compare \modelname~with SecureML to test its efficiency. 
To do this, we set the worker number to 10, use all the features, and vary network bandwidth to study the running time of both \modelname~and SecureML for passing the data once (one epoch). We show the results in Figure \ref{fig-comp1} and Figure \ref{fig-comp2}, where we set batch size to 1,024 and 4,096, respectively. 
From them, we observe that, \modelname~consistently achieves better efficiency than SecureML, especially when the network bandwidth is limited. 
Specifically, take Figure \ref{fig-comp1} for example, \modelname~improves the running speed of SecureML by 40x, 25x, 18x, 13x, in average, when the network bandwidth are 10Mbps, 20Mbps, 30Mbps, and 40Mbps, respectively. 
Moreover, we also find that, increasing batch size will also improve the speedup of \modelname~against SecureML. Take bandwidth=10Mbps for example, \modelname~improves the running speed of SecureML by 40x and 130x when batch size are 1,024 and 4,096, respectively. 
One should also notice that, for SecureML, we only count the online running time, and it will take much longer time if we count the offline Beaver's triples generation time.

\subsection{Parameter Analysis}
To further study the efficiency of \modelname, we change the number of workers, the number of features, batch size, and bandwidth, and report the running time of \modelname~per epoch. 

\nosection{Effect of worker number} 
We first use all the features, fix batch size to 8,192, and bandwidth to 32Mbps to study the effect of worker number on \modelname. 
From Figure \ref{fig-param-effect} (a), we can find that worker number significantly affects the efficiency of \modelname. With more workers, \modelname~can scale to large datasets, which indicates the scalability of our distributed implementation. 

\nosection{Effect of feature number} 
We then fix worker number to 10, batch size to 8,192, and bandwidth to 32Mbps to study the effect of feature number on \modelname. We can find from Figure \ref{fig-param-effect} (b) that \modelname~scales linearly with feature size, the same as we have analyzed in Section \ref{sec-model-lr}, which proves the scalability of \modelname. 

\nosection{Effect of batch size} 
Next, we fix worker number to 10, use all the features, and set bandwidth to 32Mbps to study the effect of batch size on \modelname. 
From Figure \ref{fig-param-effect} (c), we find that the running time of \modelname~decreases when batch size increases. This is because in each epoch, the communication complexity between $\mathcal{A}$ and $\mathcal{B}$ is $O(7n+2nd/|\textbf{B}|)$, as we analyzed in Section \ref{sec-model-lr}, and therefore, increasing batch size will decrease the running time of each epoch. 

\nosection{Effect of network bandwidth} 
Finally, we fix worker number to 10, use all the features, and set batch size to 8,192 to study the effect of bandwidth on \modelname. We observe from Figure \ref{fig-param-effect} (d) that the running time of \modelname~also decreases with the increases of bandwidth, which is consistent with common sense. 
However, when bandwidth is large enough, the running time of \modelname~tends to be stable, this is because computation time, instead of communication time, becomes the new bottleneck.



\section{Conclusion and Future Work}
In this paper, to solve the efficiency and security problem of the existing secure and privacy-preserving logistic regression models, we propose \modelname, which combines homomorphic encryption and secret sharing to build seCure lArge-scalE SpArse logistic Regression model. 
We then implemented \modelname~distributedly across different parties. Finally, we deployed \modelname~into a risk control task and conducted experiments on it. 
In future, we plan to customize \modelname~for more machine learning models and deploy them for more applications. 

\balance
\bibliographystyle{ACM-Reference-Format}
\bibliography{reference}


\begin{thebibliography}{50}


\ifx \showCODEN    \undefined \def \showCODEN     #1{\unskip}     \fi
\ifx \showDOI      \undefined \def \showDOI       #1{#1}\fi
\ifx \showISBNx    \undefined \def \showISBNx     #1{\unskip}     \fi
\ifx \showISBNxiii \undefined \def \showISBNxiii  #1{\unskip}     \fi
\ifx \showISSN     \undefined \def \showISSN      #1{\unskip}     \fi
\ifx \showLCCN     \undefined \def \showLCCN      #1{\unskip}     \fi
\ifx \shownote     \undefined \def \shownote      #1{#1}          \fi
\ifx \showarticletitle \undefined \def \showarticletitle #1{#1}   \fi
\ifx \showURL      \undefined \def \showURL       {\relax}        \fi
\providecommand\bibfield[2]{#2}
\providecommand\bibinfo[2]{#2}
\providecommand\natexlab[1]{#1}
\providecommand\showeprint[2][]{arXiv:#2}

\bibitem[\protect\citeauthoryear{Acar, Aksu, Uluagac, and Conti}{Acar
  et~al\mbox{.}}{2018}]%
        {acar2018survey}
\bibfield{author}{\bibinfo{person}{Abbas Acar}, \bibinfo{person}{Hidayet Aksu},
  \bibinfo{person}{A~Selcuk Uluagac}, {and} \bibinfo{person}{Mauro Conti}.}
  \bibinfo{year}{2018}\natexlab{}.
\newblock \showarticletitle{A survey on homomorphic encryption schemes: Theory
  and implementation}.
\newblock \bibinfo{journal}{\emph{CSUR}} \bibinfo{volume}{51},
  \bibinfo{number}{4} (\bibinfo{year}{2018}), \bibinfo{pages}{79}.
\newblock


\bibitem[\protect\citeauthoryear{Agrawal, Shahin~Shamsabadi, Kusner, and
  Gasc{\'o}n}{Agrawal et~al\mbox{.}}{2019}]%
        {agrawal2019quotient}
\bibfield{author}{\bibinfo{person}{Nitin Agrawal}, \bibinfo{person}{Ali
  Shahin~Shamsabadi}, \bibinfo{person}{Matt~J Kusner}, {and}
  \bibinfo{person}{Adri{\`a} Gasc{\'o}n}.} \bibinfo{year}{2019}\natexlab{}.
\newblock \showarticletitle{QUOTIENT: Two-Party Secure Neural Network Training
  and Prediction}. In \bibinfo{booktitle}{\emph{CCS}}. ACM,
  \bibinfo{pages}{1231--1247}.
\newblock


\bibitem[\protect\citeauthoryear{Aono, Hayashi, Trieu~Phong, and Wang}{Aono
  et~al\mbox{.}}{2016}]%
        {aono2016scalable}
\bibfield{author}{\bibinfo{person}{Yoshinori Aono}, \bibinfo{person}{Takuya
  Hayashi}, \bibinfo{person}{Le Trieu~Phong}, {and} \bibinfo{person}{Lihua
  Wang}.} \bibinfo{year}{2016}\natexlab{}.
\newblock \showarticletitle{Scalable and secure logistic regression via
  homomorphic encryption}. In \bibinfo{booktitle}{\emph{CODASPY}}. ACM,
  \bibinfo{pages}{142--144}.
\newblock


\bibitem[\protect\citeauthoryear{Badawi, Chao, Lin, Mun, Jie, Tan, Nan, Aung,
  and Chandrasekhar}{Badawi et~al\mbox{.}}{2018}]%
        {badawi2018alexnet}
\bibfield{author}{\bibinfo{person}{Ahmad~Al Badawi}, \bibinfo{person}{Jin
  Chao}, \bibinfo{person}{Jie Lin}, \bibinfo{person}{Chan~Fook Mun},
  \bibinfo{person}{Sim~Jun Jie}, \bibinfo{person}{Benjamin Hong~Meng Tan},
  \bibinfo{person}{Xiao Nan}, \bibinfo{person}{Khin Mi~Mi Aung}, {and}
  \bibinfo{person}{Vijay~Ramaseshan Chandrasekhar}.}
  \bibinfo{year}{2018}\natexlab{}.
\newblock \showarticletitle{The AlexNet moment for homomorphic encryption:
  HCNN, the first homomorphic CNN on encrypted data with GPUs}.
\newblock \bibinfo{journal}{\emph{arXiv preprint arXiv:1811.00778}}
  (\bibinfo{year}{2018}).
\newblock


\bibitem[\protect\citeauthoryear{Beaver}{Beaver}{1991}]%
        {beaver1991efficient}
\bibfield{author}{\bibinfo{person}{Donald Beaver}.}
  \bibinfo{year}{1991}\natexlab{}.
\newblock \showarticletitle{Efficient multiparty protocols using circuit
  randomization}. In \bibinfo{booktitle}{\emph{Cryptology}}. Springer,
  \bibinfo{pages}{420--432}.
\newblock


\bibitem[\protect\citeauthoryear{Boyle, Gilboa, and Ishai}{Boyle
  et~al\mbox{.}}{2015}]%
        {boyle2015function}
\bibfield{author}{\bibinfo{person}{Elette Boyle}, \bibinfo{person}{Niv Gilboa},
  {and} \bibinfo{person}{Yuval Ishai}.} \bibinfo{year}{2015}\natexlab{}.
\newblock \showarticletitle{Function secret sharing}. In
  \bibinfo{booktitle}{\emph{Eurocrypt}}. Springer, \bibinfo{pages}{337--367}.
\newblock


\bibitem[\protect\citeauthoryear{Chen, Li, Fang, Zhou, Wang, Wang, Yang, Liu,
  and Wang}{Chen et~al\mbox{.}}{2020a}]%
        {chen2020secret}
\bibfield{author}{\bibinfo{person}{Chaochao Chen}, \bibinfo{person}{Liang Li},
  \bibinfo{person}{Wenjing Fang}, \bibinfo{person}{Jun Zhou},
  \bibinfo{person}{Li Wang}, \bibinfo{person}{Lei Wang},
  \bibinfo{person}{Shuang Yang}, \bibinfo{person}{Alex Liu}, {and}
  \bibinfo{person}{Hao Wang}.} \bibinfo{year}{2020}\natexlab{a}.
\newblock \showarticletitle{Secret Sharing based Secure Regressions with
  Applications}.
\newblock \bibinfo{journal}{\emph{arXiv preprint arXiv:2004.04898}}
  (\bibinfo{year}{2020}).
\newblock


\bibitem[\protect\citeauthoryear{Chen, Li, Wu, Hong, Wang, and Zhou}{Chen
  et~al\mbox{.}}{2020b}]%
        {chen2020secure}
\bibfield{author}{\bibinfo{person}{Chaochao Chen}, \bibinfo{person}{Liang Li},
  \bibinfo{person}{Bingzhe Wu}, \bibinfo{person}{Cheng Hong},
  \bibinfo{person}{Li Wang}, {and} \bibinfo{person}{Jun Zhou}.}
  \bibinfo{year}{2020}\natexlab{b}.
\newblock \showarticletitle{Secure social recommendation based on secret
  sharing}. In \bibinfo{booktitle}{\emph{ECAI}}. \bibinfo{pages}{506--512}.
\newblock


\bibitem[\protect\citeauthoryear{Chen, Gilad-Bachrach, Han, Huang, Jalali,
  Laine, and Lauter}{Chen et~al\mbox{.}}{2018}]%
        {chen2018logistic}
\bibfield{author}{\bibinfo{person}{Hao Chen}, \bibinfo{person}{Ran
  Gilad-Bachrach}, \bibinfo{person}{Kyoohyung Han}, \bibinfo{person}{Zhicong
  Huang}, \bibinfo{person}{Amir Jalali}, \bibinfo{person}{Kim Laine}, {and}
  \bibinfo{person}{Kristin Lauter}.} \bibinfo{year}{2018}\natexlab{}.
\newblock \showarticletitle{Logistic regression over encrypted data from fully
  homomorphic encryption}.
\newblock \bibinfo{journal}{\emph{BMC medical genomics}} \bibinfo{volume}{11},
  \bibinfo{number}{4} (\bibinfo{year}{2018}), \bibinfo{pages}{81}.
\newblock


\bibitem[\protect\citeauthoryear{Chen and Asch}{Chen and Asch}{2017}]%
        {chen2017machine}
\bibfield{author}{\bibinfo{person}{Jonathan~H Chen} {and}
  \bibinfo{person}{Steven~M Asch}.} \bibinfo{year}{2017}\natexlab{}.
\newblock \showarticletitle{Machine learning and prediction in
  medicine—beyond the peak of inflated expectations}.
\newblock \bibinfo{journal}{\emph{The New England journal of medicine}}
  \bibinfo{volume}{376}, \bibinfo{number}{26} (\bibinfo{year}{2017}),
  \bibinfo{pages}{2507}.
\newblock


\bibitem[\protect\citeauthoryear{Chen, Pastro, and Raykova}{Chen
  et~al\mbox{.}}{2019}]%
        {chen2019secure}
\bibfield{author}{\bibinfo{person}{Valerie Chen}, \bibinfo{person}{Valerio
  Pastro}, {and} \bibinfo{person}{Mariana Raykova}.}
  \bibinfo{year}{2019}\natexlab{}.
\newblock \showarticletitle{Secure computation for machine learning with SPDZ}.
\newblock \bibinfo{journal}{\emph{arXiv preprint arXiv:1901.00329}}
  (\bibinfo{year}{2019}).
\newblock


\bibitem[\protect\citeauthoryear{Damg{\aa}rd, Geisler, Kr{\o}igaard, and
  Nielsen}{Damg{\aa}rd et~al\mbox{.}}{2009}]%
        {damgaard2009asynchronous}
\bibfield{author}{\bibinfo{person}{Ivan Damg{\aa}rd}, \bibinfo{person}{Martin
  Geisler}, \bibinfo{person}{Mikkel Kr{\o}igaard}, {and}
  \bibinfo{person}{Jesper~Buus Nielsen}.} \bibinfo{year}{2009}\natexlab{}.
\newblock \showarticletitle{Asynchronous multiparty computation: Theory and
  implementation}. In \bibinfo{booktitle}{\emph{International Workshop on
  Public Key Cryptography}}. Springer, \bibinfo{pages}{160--179}.
\newblock


\bibitem[\protect\citeauthoryear{Damg{\aa}rd, Pastro, Smart, and
  Zakarias}{Damg{\aa}rd et~al\mbox{.}}{2012}]%
        {damgaard2012multiparty}
\bibfield{author}{\bibinfo{person}{Ivan Damg{\aa}rd}, \bibinfo{person}{Valerio
  Pastro}, \bibinfo{person}{Nigel Smart}, {and} \bibinfo{person}{Sarah
  Zakarias}.} \bibinfo{year}{2012}\natexlab{}.
\newblock \showarticletitle{Multiparty computation from somewhat homomorphic
  encryption}. In \bibinfo{booktitle}{\emph{Cryptology}}. Springer,
  \bibinfo{pages}{643--662}.
\newblock


\bibitem[\protect\citeauthoryear{De~Cock, Dowsley, Horst, Katti, Nascimento,
  Poon, and Truex}{De~Cock et~al\mbox{.}}{2017}]%
        {de2017efficient}
\bibfield{author}{\bibinfo{person}{Martine De~Cock}, \bibinfo{person}{Rafael
  Dowsley}, \bibinfo{person}{Caleb Horst}, \bibinfo{person}{Raj Katti},
  \bibinfo{person}{Anderson~CA Nascimento}, \bibinfo{person}{Wing-Sea Poon},
  {and} \bibinfo{person}{Stacey Truex}.} \bibinfo{year}{2017}\natexlab{}.
\newblock \showarticletitle{Efficient and private scoring of decision trees,
  support vector machines and logistic regression models based on
  pre-computation}.
\newblock \bibinfo{journal}{\emph{TDSC}} \bibinfo{volume}{16},
  \bibinfo{number}{2} (\bibinfo{year}{2017}), \bibinfo{pages}{217--230}.
\newblock


\bibitem[\protect\citeauthoryear{Demmler, Schneider, and Zohner}{Demmler
  et~al\mbox{.}}{2015}]%
        {demmler2015aby}
\bibfield{author}{\bibinfo{person}{Daniel Demmler}, \bibinfo{person}{Thomas
  Schneider}, {and} \bibinfo{person}{Michael Zohner}.}
  \bibinfo{year}{2015}\natexlab{}.
\newblock \showarticletitle{ABY-A Framework for Efficient Mixed-Protocol Secure
  Two-Party Computation.}. In \bibinfo{booktitle}{\emph{NDSS}}.
\newblock


\bibitem[\protect\citeauthoryear{Esperan{\c{c}}a, Aslett, and
  Holmes}{Esperan{\c{c}}a et~al\mbox{.}}{2017}]%
        {esperancca2017encrypted}
\bibfield{author}{\bibinfo{person}{Pedro~M Esperan{\c{c}}a},
  \bibinfo{person}{Louis~JM Aslett}, {and} \bibinfo{person}{Chris~C Holmes}.}
  \bibinfo{year}{2017}\natexlab{}.
\newblock \showarticletitle{Encrypted accelerated least squares regression}.
\newblock \bibinfo{journal}{\emph{arXiv preprint arXiv:1703.00839}}
  (\bibinfo{year}{2017}).
\newblock


\bibitem[\protect\citeauthoryear{Fang, Chen, Tan, Yu, Lu, Wang, Wang, Zhou,
  et~al\mbox{.}}{Fang et~al\mbox{.}}{2020}]%
        {fang2020hybrid}
\bibfield{author}{\bibinfo{person}{Wenjing Fang}, \bibinfo{person}{Chaochao
  Chen}, \bibinfo{person}{Jin Tan}, \bibinfo{person}{Chaofan Yu},
  \bibinfo{person}{Yufei Lu}, \bibinfo{person}{Li Wang}, \bibinfo{person}{Lei
  Wang}, \bibinfo{person}{Jun Zhou}, {et~al\mbox{.}}}
  \bibinfo{year}{2020}\natexlab{}.
\newblock \showarticletitle{A Hybrid-Domain Framework for Secure Gradient Tree
  Boosting}.
\newblock \bibinfo{journal}{\emph{arXiv preprint arXiv:2005.08479}}
  (\bibinfo{year}{2020}).
\newblock


\bibitem[\protect\citeauthoryear{Gasc{\'o}n, Schoppmann, Balle, Raykova,
  Doerner, Zahur, and Evans}{Gasc{\'o}n et~al\mbox{.}}{2017}]%
        {gascon2017privacy}
\bibfield{author}{\bibinfo{person}{Adri{\`a} Gasc{\'o}n},
  \bibinfo{person}{Phillipp Schoppmann}, \bibinfo{person}{Borja Balle},
  \bibinfo{person}{Mariana Raykova}, \bibinfo{person}{Jack Doerner},
  \bibinfo{person}{Samee Zahur}, {and} \bibinfo{person}{David Evans}.}
  \bibinfo{year}{2017}\natexlab{}.
\newblock \showarticletitle{Privacy-preserving distributed linear regression on
  high-dimensional data}.
\newblock \bibinfo{journal}{\emph{PETs}}, \bibinfo{pages}{345--364}.
\newblock


\bibitem[\protect\citeauthoryear{Gilad-Bachrach, Dowlin, Laine, Lauter,
  Naehrig, and Wernsing}{Gilad-Bachrach et~al\mbox{.}}{2016}]%
        {gilad2016cryptonets}
\bibfield{author}{\bibinfo{person}{Ran Gilad-Bachrach}, \bibinfo{person}{Nathan
  Dowlin}, \bibinfo{person}{Kim Laine}, \bibinfo{person}{Kristin Lauter},
  \bibinfo{person}{Michael Naehrig}, {and} \bibinfo{person}{John Wernsing}.}
  \bibinfo{year}{2016}\natexlab{}.
\newblock \showarticletitle{Cryptonets: Applying neural networks to encrypted
  data with high throughput and accuracy}. In \bibinfo{booktitle}{\emph{ICML}}.
  \bibinfo{pages}{201--210}.
\newblock


\bibitem[\protect\citeauthoryear{Goldreich}{Goldreich}{2009}]%
        {goldreich2009foundations}
\bibfield{author}{\bibinfo{person}{Oded Goldreich}.}
  \bibinfo{year}{2009}\natexlab{}.
\newblock \bibinfo{booktitle}{\emph{Foundations of cryptography: volume 2,
  basic applications}}.
\newblock \bibinfo{publisher}{Cambridge university press}.
\newblock


\bibitem[\protect\citeauthoryear{Hall, Fienberg, and Nardi}{Hall
  et~al\mbox{.}}{2011}]%
        {hall2011secure}
\bibfield{author}{\bibinfo{person}{Rob Hall}, \bibinfo{person}{Stephen~E
  Fienberg}, {and} \bibinfo{person}{Yuval Nardi}.}
  \bibinfo{year}{2011}\natexlab{}.
\newblock \showarticletitle{Secure multiple linear regression based on
  homomorphic encryption}.
\newblock \bibinfo{journal}{\emph{Journal of Official Statistics}}
  \bibinfo{volume}{27}, \bibinfo{number}{4} (\bibinfo{year}{2011}),
  \bibinfo{pages}{669}.
\newblock


\bibitem[\protect\citeauthoryear{Han, Hong, Cheon, and Park}{Han
  et~al\mbox{.}}{2019}]%
        {Han2019logistic}
\bibfield{author}{\bibinfo{person}{Kyoohyung Han}, \bibinfo{person}{Seungwan
  Hong}, \bibinfo{person}{Jung~Hee Cheon}, {and} \bibinfo{person}{Daejun
  Park}.} \bibinfo{year}{2019}\natexlab{}.
\newblock \showarticletitle{Logistic Regression on Homomorphic Encrypted Data
  at Scale}. In \bibinfo{booktitle}{\emph{IAAI}}. \bibinfo{pages}{9466--9471}.
\newblock
\urldef\tempurl%
\url{https://doi.org/10.1609/aaai.v33i01.33019466}
\showDOI{\tempurl}


\bibitem[\protect\citeauthoryear{Hardy, Henecka, Ivey-Law, Nock, Patrini,
  Smith, and Thorne}{Hardy et~al\mbox{.}}{2017}]%
        {hardy2017private}
\bibfield{author}{\bibinfo{person}{Stephen Hardy}, \bibinfo{person}{Wilko
  Henecka}, \bibinfo{person}{Hamish Ivey-Law}, \bibinfo{person}{Richard Nock},
  \bibinfo{person}{Giorgio Patrini}, \bibinfo{person}{Guillaume Smith}, {and}
  \bibinfo{person}{Brian Thorne}.} \bibinfo{year}{2017}\natexlab{}.
\newblock \showarticletitle{Private federated learning on vertically
  partitioned data via entity resolution and additively homomorphic
  encryption}.
\newblock \bibinfo{journal}{\emph{arXiv preprint arXiv:1711.10677}}
  (\bibinfo{year}{2017}).
\newblock


\bibitem[\protect\citeauthoryear{Hazay and Lindell}{Hazay and Lindell}{2010}]%
        {hazay2010efficient}
\bibfield{author}{\bibinfo{person}{Carmit Hazay} {and} \bibinfo{person}{Yehuda
  Lindell}.} \bibinfo{year}{2010}\natexlab{}.
\newblock \bibinfo{booktitle}{\emph{Efficient secure two-party protocols:
  Techniques and constructions}}.
\newblock \bibinfo{publisher}{Springer Science \& Business Media}.
\newblock


\bibitem[\protect\citeauthoryear{Hesamifard, Takabi, and Ghasemi}{Hesamifard
  et~al\mbox{.}}{2017}]%
        {hesamifard2017cryptodl}
\bibfield{author}{\bibinfo{person}{Ehsan Hesamifard}, \bibinfo{person}{Hassan
  Takabi}, {and} \bibinfo{person}{Mehdi Ghasemi}.}
  \bibinfo{year}{2017}\natexlab{}.
\newblock \showarticletitle{Cryptodl: Deep neural networks over encrypted
  data}.
\newblock \bibinfo{journal}{\emph{arXiv preprint arXiv:1711.05189}}
  (\bibinfo{year}{2017}).
\newblock


\bibitem[\protect\citeauthoryear{Jiang, Hamer, Wang, Jiang, Kim, Song, Xia,
  Mohammed, Sadat, and Wang}{Jiang et~al\mbox{.}}{2018}]%
        {jiang2018securelr}
\bibfield{author}{\bibinfo{person}{Yichen Jiang}, \bibinfo{person}{Jenny
  Hamer}, \bibinfo{person}{Chenghong Wang}, \bibinfo{person}{Xiaoqian Jiang},
  \bibinfo{person}{Miran Kim}, \bibinfo{person}{Yongsoo Song},
  \bibinfo{person}{Yuhou Xia}, \bibinfo{person}{Noman Mohammed},
  \bibinfo{person}{Md~Nazmus Sadat}, {and} \bibinfo{person}{Shuang Wang}.}
  \bibinfo{year}{2018}\natexlab{}.
\newblock \showarticletitle{SecureLR: Secure logistic regression model via a
  hybrid cryptographic protocol}.
\newblock \bibinfo{journal}{\emph{IEEE/ACM TCBB}} \bibinfo{volume}{16},
  \bibinfo{number}{1} (\bibinfo{year}{2018}), \bibinfo{pages}{113--123}.
\newblock


\bibitem[\protect\citeauthoryear{Juvekar, Vaikuntanathan, and
  Chandrakasan}{Juvekar et~al\mbox{.}}{2018}]%
        {juvekar2018gazelle}
\bibfield{author}{\bibinfo{person}{Chiraag Juvekar}, \bibinfo{person}{Vinod
  Vaikuntanathan}, {and} \bibinfo{person}{Anantha Chandrakasan}.}
  \bibinfo{year}{2018}\natexlab{}.
\newblock \showarticletitle{$\{$GAZELLE$\}$: A Low Latency Framework for Secure
  Neural Network Inference}. In \bibinfo{booktitle}{\emph{USENIX Security}}.
  \bibinfo{pages}{1651--1669}.
\newblock


\bibitem[\protect\citeauthoryear{Kim, Song, Wang, Xia, and Jiang}{Kim
  et~al\mbox{.}}{2018}]%
        {kim2018secure}
\bibfield{author}{\bibinfo{person}{Miran Kim}, \bibinfo{person}{Yongsoo Song},
  \bibinfo{person}{Shuang Wang}, \bibinfo{person}{Yuhou Xia}, {and}
  \bibinfo{person}{Xiaoqian Jiang}.} \bibinfo{year}{2018}\natexlab{}.
\newblock \showarticletitle{Secure logistic regression based on homomorphic
  encryption: Design and evaluation}.
\newblock \bibinfo{journal}{\emph{JMIR medical informatics}}
  \bibinfo{volume}{6}, \bibinfo{number}{2} (\bibinfo{year}{2018}),
  \bibinfo{pages}{e19}.
\newblock


\bibitem[\protect\citeauthoryear{Li, Andersen, Park, Smola, Ahmed, Josifovski,
  Long, Shekita, and Su}{Li et~al\mbox{.}}{2014}]%
        {li2014scaling}
\bibfield{author}{\bibinfo{person}{Mu Li}, \bibinfo{person}{David~G Andersen},
  \bibinfo{person}{Jun~Woo Park}, \bibinfo{person}{Alexander~J Smola},
  \bibinfo{person}{Amr Ahmed}, \bibinfo{person}{Vanja Josifovski},
  \bibinfo{person}{James Long}, \bibinfo{person}{Eugene~J Shekita}, {and}
  \bibinfo{person}{Bor-Yiing Su}.} \bibinfo{year}{2014}\natexlab{}.
\newblock \showarticletitle{Scaling distributed machine learning with the
  parameter server}. In \bibinfo{booktitle}{\emph{USENIX Security}}.
  \bibinfo{pages}{583--598}.
\newblock


\bibitem[\protect\citeauthoryear{Li, Duan, Yu, Zhao, and Xu}{Li
  et~al\mbox{.}}{2018}]%
        {li2018privpy}
\bibfield{author}{\bibinfo{person}{Yi Li}, \bibinfo{person}{Yitao Duan},
  \bibinfo{person}{Yu Yu}, \bibinfo{person}{Shuoyao Zhao}, {and}
  \bibinfo{person}{Wei Xu}.} \bibinfo{year}{2018}\natexlab{}.
\newblock \showarticletitle{PrivPy: Enabling Scalable and General
  Privacy-Preserving Machine Learning}.
\newblock \bibinfo{journal}{\emph{arXiv preprint arXiv:1801.10117}}
  (\bibinfo{year}{2018}).
\newblock


\bibitem[\protect\citeauthoryear{Li, Huang, Chen, and Hong}{Li
  et~al\mbox{.}}{2019}]%
        {li2019quantification}
\bibfield{author}{\bibinfo{person}{Zhaorui Li}, \bibinfo{person}{Zhicong
  Huang}, \bibinfo{person}{Chaochao Chen}, {and} \bibinfo{person}{Cheng Hong}.}
  \bibinfo{year}{2019}\natexlab{}.
\newblock \showarticletitle{Quantification of the Leakage in Federated
  Learning}.
\newblock \bibinfo{journal}{\emph{arXiv preprint arXiv:1910.05467}}
  (\bibinfo{year}{2019}).
\newblock


\bibitem[\protect\citeauthoryear{Mohassel and Rindal}{Mohassel and
  Rindal}{2018}]%
        {mohassel2018aby}
\bibfield{author}{\bibinfo{person}{Payman Mohassel} {and}
  \bibinfo{person}{Peter Rindal}.} \bibinfo{year}{2018}\natexlab{}.
\newblock \showarticletitle{ABY 3: a mixed protocol framework for machine
  learning}. In \bibinfo{booktitle}{\emph{CCS}}. ACM, \bibinfo{pages}{35--52}.
\newblock


\bibitem[\protect\citeauthoryear{Mohassel and Zhang}{Mohassel and
  Zhang}{2017}]%
        {mohassel2017secureml}
\bibfield{author}{\bibinfo{person}{Payman Mohassel} {and}
  \bibinfo{person}{Yupeng Zhang}.} \bibinfo{year}{2017}\natexlab{}.
\newblock \showarticletitle{Secureml: A system for scalable privacy-preserving
  machine learning}. In \bibinfo{booktitle}{\emph{S\&P}}. IEEE,
  \bibinfo{pages}{19--38}.
\newblock


\bibitem[\protect\citeauthoryear{Okamoto and Uchiyama}{Okamoto and
  Uchiyama}{1998}]%
        {okamoto1998new}
\bibfield{author}{\bibinfo{person}{Tatsuaki Okamoto} {and}
  \bibinfo{person}{Shigenori Uchiyama}.} \bibinfo{year}{1998}\natexlab{}.
\newblock \showarticletitle{A new public-key cryptosystem as secure as
  factoring}. In \bibinfo{booktitle}{\emph{Eurocrypt}}. Springer,
  \bibinfo{pages}{308--318}.
\newblock


\bibitem[\protect\citeauthoryear{Paillier}{Paillier}{1999}]%
        {paillier1999public}
\bibfield{author}{\bibinfo{person}{Pascal Paillier}.}
  \bibinfo{year}{1999}\natexlab{}.
\newblock \showarticletitle{Public-key cryptosystems based on composite degree
  residuosity classes}. In \bibinfo{booktitle}{\emph{Eurocrypt}}. Springer,
  \bibinfo{pages}{223--238}.
\newblock


\bibitem[\protect\citeauthoryear{Pinkas, Schneider, and Zohner}{Pinkas
  et~al\mbox{.}}{2014}]%
        {pinkas2014faster}
\bibfield{author}{\bibinfo{person}{Benny Pinkas}, \bibinfo{person}{Thomas
  Schneider}, {and} \bibinfo{person}{Michael Zohner}.}
  \bibinfo{year}{2014}\natexlab{}.
\newblock \showarticletitle{Faster Private Set Intersection Based on $\{$OT$\}$
  Extension}. In \bibinfo{booktitle}{\emph{USENIX Security}}.
  \bibinfo{pages}{797--812}.
\newblock


\bibitem[\protect\citeauthoryear{Rouhani, Riazi, and Koushanfar}{Rouhani
  et~al\mbox{.}}{2018}]%
        {rouhani2018deepsecure}
\bibfield{author}{\bibinfo{person}{Bita~Darvish Rouhani},
  \bibinfo{person}{M~Sadegh Riazi}, {and} \bibinfo{person}{Farinaz
  Koushanfar}.} \bibinfo{year}{2018}\natexlab{}.
\newblock \showarticletitle{Deepsecure: Scalable provably-secure deep
  learning}. In \bibinfo{booktitle}{\emph{DAC}}. ACM, \bibinfo{pages}{2}.
\newblock


\bibitem[\protect\citeauthoryear{Schoppmann, Gasc{\'o}n, Raykova, and
  Pinkas}{Schoppmann et~al\mbox{.}}{2019}]%
        {schoppmann2019make}
\bibfield{author}{\bibinfo{person}{Phillipp Schoppmann},
  \bibinfo{person}{Adri{\`a} Gasc{\'o}n}, \bibinfo{person}{Mariana Raykova},
  {and} \bibinfo{person}{Benny Pinkas}.} \bibinfo{year}{2019}\natexlab{}.
\newblock \showarticletitle{Make some room for the zeros: Data sparsity in
  secure distributed machine learning}. In \bibinfo{booktitle}{\emph{CCS}}.
  \bibinfo{pages}{1335--1350}.
\newblock


\bibitem[\protect\citeauthoryear{Shamir}{Shamir}{1979}]%
        {shamir1979share}
\bibfield{author}{\bibinfo{person}{Adi Shamir}.}
  \bibinfo{year}{1979}\natexlab{}.
\newblock \showarticletitle{How to share a secret}.
\newblock \bibinfo{journal}{\emph{Commun. ACM}} \bibinfo{volume}{22},
  \bibinfo{number}{11} (\bibinfo{year}{1979}), \bibinfo{pages}{612--613}.
\newblock


\bibitem[\protect\citeauthoryear{Shi, Jiang, Dai, Jiang, Tang, Ohno-Machado,
  and Wang}{Shi et~al\mbox{.}}{2016}]%
        {shi2016secure}
\bibfield{author}{\bibinfo{person}{Haoyi Shi}, \bibinfo{person}{Chao Jiang},
  \bibinfo{person}{Wenrui Dai}, \bibinfo{person}{Xiaoqian Jiang},
  \bibinfo{person}{Yuzhe Tang}, \bibinfo{person}{Lucila Ohno-Machado}, {and}
  \bibinfo{person}{Shuang Wang}.} \bibinfo{year}{2016}\natexlab{}.
\newblock \showarticletitle{Secure multi-pArty computation grid LOgistic
  REgression (SMAC-GLORE)}.
\newblock \bibinfo{journal}{\emph{BMC medical informatics and decision making}}
  \bibinfo{volume}{16}, \bibinfo{number}{3} (\bibinfo{year}{2016}),
  \bibinfo{pages}{89}.
\newblock


\bibitem[\protect\citeauthoryear{Wagh, Gupta, and Chandran}{Wagh
  et~al\mbox{.}}{2018}]%
        {wagh2018securenn}
\bibfield{author}{\bibinfo{person}{Sameer Wagh}, \bibinfo{person}{Divya Gupta},
  {and} \bibinfo{person}{Nishanth Chandran}.} \bibinfo{year}{2018}\natexlab{}.
\newblock \showarticletitle{SecureNN: Efficient and Private Neural Network
  Training.}
\newblock \bibinfo{journal}{\emph{IACR Cryptology ePrint Archive}}
  \bibinfo{volume}{2018} (\bibinfo{year}{2018}), \bibinfo{pages}{442}.
\newblock


\bibitem[\protect\citeauthoryear{Wu, Teruya, Kawamoto, Sakuma, and Kikuchi}{Wu
  et~al\mbox{.}}{2013}]%
        {wu2013privacy}
\bibfield{author}{\bibinfo{person}{Shuang Wu}, \bibinfo{person}{Tadanori
  Teruya}, \bibinfo{person}{Junpei Kawamoto}, \bibinfo{person}{Jun Sakuma},
  {and} \bibinfo{person}{Hiroaki Kikuchi}.} \bibinfo{year}{2013}\natexlab{}.
\newblock \showarticletitle{Privacy-preservation for stochastic gradient
  descent application to secure logistic regression}. In
  \bibinfo{booktitle}{\emph{The 27th Annual Conference of the Japanese Society
  for Artificial Intelligence}}, Vol.~\bibinfo{volume}{27}.
  \bibinfo{pages}{1--4}.
\newblock


\bibitem[\protect\citeauthoryear{Xie, Wang, Boker, and Brown}{Xie
  et~al\mbox{.}}{2016}]%
        {xie2016privlogit}
\bibfield{author}{\bibinfo{person}{Wei Xie}, \bibinfo{person}{Yang Wang},
  \bibinfo{person}{Steven~M Boker}, {and} \bibinfo{person}{Donald~E Brown}.}
  \bibinfo{year}{2016}\natexlab{}.
\newblock \showarticletitle{Privlogit: Efficient privacy-preserving logistic
  regression by tailoring numerical optimizers}.
\newblock \bibinfo{journal}{\emph{arXiv preprint arXiv:1611.01170}}
  (\bibinfo{year}{2016}).
\newblock


\bibitem[\protect\citeauthoryear{Xu, Joshi, and Li}{Xu et~al\mbox{.}}{2019}]%
        {xu2019cryptonn}
\bibfield{author}{\bibinfo{person}{Runhua Xu}, \bibinfo{person}{James~BD
  Joshi}, {and} \bibinfo{person}{Chao Li}.} \bibinfo{year}{2019}\natexlab{}.
\newblock \showarticletitle{CryptoNN: Training Neural Networks over Encrypted
  Data}.
\newblock \bibinfo{journal}{\emph{arXiv preprint arXiv:1904.07303}}
  (\bibinfo{year}{2019}).
\newblock


\bibitem[\protect\citeauthoryear{Yang, Liu, Chen, and Tong}{Yang
  et~al\mbox{.}}{2019}]%
        {yang2019federated}
\bibfield{author}{\bibinfo{person}{Qiang Yang}, \bibinfo{person}{Yang Liu},
  \bibinfo{person}{Tianjian Chen}, {and} \bibinfo{person}{Yongxin Tong}.}
  \bibinfo{year}{2019}\natexlab{}.
\newblock \showarticletitle{Federated machine learning: Concept and
  applications}.
\newblock \bibinfo{journal}{\emph{TIST}} \bibinfo{volume}{10},
  \bibinfo{number}{2} (\bibinfo{year}{2019}), \bibinfo{pages}{12}.
\newblock


\bibitem[\protect\citeauthoryear{Yao}{Yao}{1982}]%
        {yao1982protocols}
\bibfield{author}{\bibinfo{person}{Andrew~C Yao}.}
  \bibinfo{year}{1982}\natexlab{}.
\newblock \showarticletitle{Protocols for secure computations}. In
  \bibinfo{booktitle}{\emph{FOCS}}. IEEE, \bibinfo{pages}{160--164}.
\newblock


\bibitem[\protect\citeauthoryear{Yuan and Yu}{Yuan and Yu}{2013}]%
        {yuan2013privacy}
\bibfield{author}{\bibinfo{person}{Jiawei Yuan} {and} \bibinfo{person}{Shucheng
  Yu}.} \bibinfo{year}{2013}\natexlab{}.
\newblock \showarticletitle{Privacy preserving back-propagation neural network
  learning made practical with cloud computing}.
\newblock \bibinfo{journal}{\emph{TPDS}} \bibinfo{volume}{25},
  \bibinfo{number}{1} (\bibinfo{year}{2013}), \bibinfo{pages}{212--221}.
\newblock


\bibitem[\protect\citeauthoryear{Zhang, Zhou, Zheng, Feng, Li, Liu, Li, Zhang,
  Chen, Li, et~al\mbox{.}}{Zhang et~al\mbox{.}}{2019}]%
        {zhang2019distributed}
\bibfield{author}{\bibinfo{person}{Ya-Lin Zhang}, \bibinfo{person}{Jun Zhou},
  \bibinfo{person}{Wenhao Zheng}, \bibinfo{person}{Ji Feng},
  \bibinfo{person}{Longfei Li}, \bibinfo{person}{Ziqi Liu},
  \bibinfo{person}{Ming Li}, \bibinfo{person}{Zhiqiang Zhang},
  \bibinfo{person}{Chaochao Chen}, \bibinfo{person}{Xiaolong Li},
  {et~al\mbox{.}}} \bibinfo{year}{2019}\natexlab{}.
\newblock \showarticletitle{Distributed Deep Forest and its Application to
  Automatic Detection of Cash-Out Fraud}.
\newblock \bibinfo{journal}{\emph{TIST}} \bibinfo{volume}{10},
  \bibinfo{number}{5} (\bibinfo{year}{2019}), \bibinfo{pages}{55}.
\newblock


\bibitem[\protect\citeauthoryear{Zheng, Chen, Liu, Wu, Wu, Wang, Wang, Zhou,
  and Yang}{Zheng et~al\mbox{.}}{2020}]%
        {zheng2020industrial}
\bibfield{author}{\bibinfo{person}{Longfei Zheng}, \bibinfo{person}{Chaochao
  Chen}, \bibinfo{person}{Yingting Liu}, \bibinfo{person}{Bingzhe Wu},
  \bibinfo{person}{Xibin Wu}, \bibinfo{person}{Li Wang}, \bibinfo{person}{Lei
  Wang}, \bibinfo{person}{Jun Zhou}, {and} \bibinfo{person}{Shuang Yang}.}
  \bibinfo{year}{2020}\natexlab{}.
\newblock \showarticletitle{Industrial scale privacy preserving deep neural
  network}.
\newblock \bibinfo{journal}{\emph{arXiv preprint arXiv:2003.05198}}
  (\bibinfo{year}{2020}).
\newblock


\bibitem[\protect\citeauthoryear{Zhou, Li, Zhao, Chen, Li, Yang, Cui, Yu, Chen,
  Ding, et~al\mbox{.}}{Zhou et~al\mbox{.}}{2017}]%
        {zhou2017kunpeng}
\bibfield{author}{\bibinfo{person}{Jun Zhou}, \bibinfo{person}{Xiaolong Li},
  \bibinfo{person}{Peilin Zhao}, \bibinfo{person}{Chaochao Chen},
  \bibinfo{person}{Longfei Li}, \bibinfo{person}{Xinxing Yang},
  \bibinfo{person}{Qing Cui}, \bibinfo{person}{Jin Yu}, \bibinfo{person}{Xu
  Chen}, \bibinfo{person}{Yi Ding}, {et~al\mbox{.}}}
  \bibinfo{year}{2017}\natexlab{}.
\newblock \showarticletitle{Kunpeng: Parameter server based distributed
  learning systems and its applications in alibaba and ant financial}. In
  \bibinfo{booktitle}{\emph{SIGKDD}}. ACM, \bibinfo{pages}{1693--1702}.
\newblock


\end{thebibliography}

\appendix

\section{Security Definition}\label{appen-a}

The two-party primitives considered in this paper pertain to the category of secure two-party computation.
Specifically, secure two-party computation is a two-party random process. It maps pairs of inputs (one from each party) to pairs of outputs (one for each party), while preserving several security properties, such as correctness, privacy, and independence of inputs \cite{hazay2010efficient}. This random process is called \emph{functionality}. Formally, denote a two-output functionality $f=(f_1,f_2)$ as $f:\{0,1\}^* \times \{0,1\}^* \rightarrow \{0,1\}^* \times \{0,1\}^*$. For a pair of inputs $(x,y)$, where $x$ is from party $P_1$ and $y$ is from party $P_2$, the output pair $(f_1(x,y),f_2(x,y))$ is a random variable. $f_1(x,y)$ is the output for $P_1$, and $f_2(x,y)$ is for $P_2$. 
During this process, neither party should learn anything more than its prescribed output.

\noindent\textbf{Definition 1} (\textit{Security in semi-honest model} \cite{goldreich2009foundations}).
Let $f = (f_1, f_2)$ be a \emph{deterministic} functionality and $\pi$ be a two-party protocol for computing $f$. Given the security parameter $\kappa$, and a pair of inputs $(x,y)$ (where $x$ is from $P_1$ and $y$ is from $P_2$), the view of $P_i$ ($i=1,2$) in the protocol $\pi$ is denoted as $\mathsf{view}^\pi_i(x, y, \kappa) = (w, r_i, m_i^1,\cdots, m_i^t)$, where $w\in\{x, y\}$, $r_i$ is the randomness used by $P_i$, and $m_i^j$ is the $j$-th message received by $P_i$; the output of $P_i$ is denoted as $\mathsf{output}_i^\pi(x, y, \kappa)$, and the joint output of the two parties is \\$\mathsf{output}^\pi(x, y, \kappa) = (\mathsf{output}_1^\pi(x, y, \kappa) ), \mathsf{output}_2^\pi(x, y, \kappa))$.
We say that $\pi$ securely computes $f$ in semi-honest model if 
\begin{itemize}[leftmargin=*] \setlength{\itemsep}{-\itemsep}
	\item There exist probabilistic polynomial-time simulators $\mathcal{S}_1$ and $\mathcal{S}_2$, such that
$$\{\mathcal{S}_1(1^\kappa,x,f_1(x,y))\}_{x,y,\kappa}\cong\{\mathsf{view}_1^\pi(x,y,\kappa)\}_{x,y,\kappa},$$
$$\{\mathcal{S}_2(1^\kappa,y,f_2(x,y))\}_{x,y,\kappa}\cong\{\mathsf{view}_2^\pi(x,y,\kappa)\}_{x,y,\kappa}.$$
	\item The joint output and the functionality output satisfy
	$$\{\mathsf{output}^\pi(x, y, \kappa)\}_{x,y,\kappa}\cong\{f(x,y)\}_{x,y,\kappa},$$
\end{itemize}
where $x, y\in \{0, 1\}^*$, and $\cong$ denotes computationally indistinguishablity.

\section{Security Proof}\label{appen-b}

\subsection{Secret Sharing in Homomorphically Encrypted Field}

\begin{small}
\begin{framed}
\textbf{Functionality of Secret Sharing in Homomorphically Encrypted Field $\mathcal{F}_{\mathsf{SSHEF}}$}\\
    \\
    \textbf{Inputs:}
    \begin{itemize}
    	\item $\mathcal{A}$ inputs a homomorphically encrypted matrix $[\![\textbf{Z}]\!]_b$ under $\mathcal{B}$'s public key $pk_b$;
        \item $\mathcal{B}$ inputs its secret key $sk_b$.
    \end{itemize}
    \textbf{Outputs:}
    \begin{itemize}
        \item $\mathcal{A}$ outputs a share $\langle\textbf{Z}\rangle_1$;
        \item $\mathcal{B}$ outputs a share $\langle\textbf{Z}\rangle_2$, where $\langle\textbf{Z}\rangle_1+\langle\textbf{Z}\rangle_2=\textbf{Z}$.
    \end{itemize}
\end{framed}
\end{small}

\paragraph{Security Proof} We show that \textbf{Protocol 2} is secure against semi-honest adversaries. Formally, we have the following theorem.

\nosection{Theorem 1}
\textit{
Assume that the additively homomorphic crypto system $\Pi=(\textsf{KeyGen}, \textsf{Enc}, \textsf{Dec})$ is indistinguishable under chosen-plaintext attacks. Then, \textbf{Protocol 2} (denoted as $\pi_2$) is secure in semi-honest model, as in \textbf{Definition 1}.
}

\begin{proof}
	We begin by proving the correctness of \textbf{Protocol 2}, \ie we prove that $\langle\textbf{Z}\rangle_1+\langle\textbf{Z}\rangle_2$ is equal to $\textbf{Z}$.
	According to the protocol execution, $\mathcal{A}$ computes $[\![\langle\textbf{Z}\rangle_2]\!]_b=[\![\textbf{Z}]\!]_b-\langle\textbf{Z}\rangle_1$.
	From the additive homomorphism of the crypto system $\Pi$, we can directly obtain the fact that the decryption of $[\![\langle\textbf{Z}\rangle_2]\!]_b$ is equal to $\textbf{Z}-\langle\textbf{Z}\rangle_1$.
	
	Therefore, it holds that $\langle\textbf{Z}\rangle_1+\langle\textbf{Z}\rangle_2=\textbf{Z}$.	
	This proves the correctness of \textbf{Protocol 2}.
	
	We now prove that we can construct two simulators $\mathcal{S}_{\mathcal{A}}$ and $\mathcal{S}_{\mathcal{B}}$, such that
	\begin{eqnarray}
		\{\mathcal{S}_{\mathcal{A}}(1^\kappa,[\![\textbf{Z}]\!]_b,\langle\textbf{Z}\rangle_1)\}_{[\![\textbf{Z}]\!]_b,sk_b,\kappa}\cong\{\mathsf{view}_{\mathcal{A}}^{\pi_2}([\![\textbf{Z}]\!]_b,sk_b,\kappa)\}_{[\![\textbf{Z}]\!]_b,sk_b,\kappa},\\
		\{\mathcal{S}_{\mathcal{B}}(1^\kappa,sk_b,\langle\textbf{Z}\rangle_2)\}_{[\![\textbf{Z}]\!]_b,sk_b,\kappa}\cong\{\mathsf{view}_{\mathcal{B}}^{\pi_2}([\![\textbf{Z}]\!]_b,sk_b,\kappa)\}_{[\![\textbf{Z}]\!]_b,sk_b,\kappa},
	\end{eqnarray}
	where $\mathsf{view}_{\mathcal{A}}^{\pi_2}$ and $\mathsf{view}_{\mathcal{B}}^{\pi_2}$ denotes the views of $\mathcal{A}$ and $\mathcal{B}$, respectively.
	
	We prove the above equations for a corrupted $\mathcal{A}$ and a corrupted $\mathcal{B}$, respectively.
	
	\paragraph{Corrupted $\mathcal{A}$} In this case, we construct a probabilistic polynomial-time simulator $\mathcal{S}_{\mathcal{A}}$ that, when given the security parameter $\kappa$, $\mathcal{A}$'s input $[\![\textbf{Z}]\!]_b$ and output $\langle\textbf{Z}\rangle_1$, can simulate the view of $\mathcal{A}$ in the protocol execution. To this end, we first analyze $\mathcal{A}$'s view $\mathsf{view}_{\mathcal{A}}^{\pi_2}([\![\textbf{Z}]\!]_b,sk_b,\kappa)$ in \textbf{Protocol 2}. In \textbf{Protocol 2}, $\mathcal{A}$ does not receive any messages from $\mathcal{B}$. Therefore, $\mathsf{view}_{\mathcal{A}}^{\pi_2}([\![\textbf{Z}]\!]_b,sk_b,\kappa)$ consists of $\mathcal{A}$'s input $[\![\textbf{Z}]\!]_b$ and the randomness $r_{\mathcal{A}}$.
	
	Given $\kappa$, $[\![\textbf{Z}]\!]_b$, and $\langle\textbf{Z}\rangle_1$, $\mathcal{S}_{\mathcal{A}}$ simply generates a simulation of $\mathsf{view}_{\mathcal{A}}^{\pi_2}([\![\textbf{Z}]\!]_b,sk_b,\kappa)$ by outputting $([\![\textbf{Z}]\!]_b, r_{\mathcal{A}})$. Therefore, we have the following two equations:
	$$\mathsf{view}_{\mathcal{A}}^{\pi_2}([\![\textbf{Z}]\!]_b,sk_b,\kappa)=([\![\textbf{Z}]\!]_b, r_{\mathcal{A}}),$$
	$$\mathcal{S}_{\mathcal{A}}(1^\kappa,[\![\textbf{Z}]\!]_b,\langle\textbf{Z}\rangle_1)=([\![\textbf{Z}]\!]_b, r_{\mathcal{A}}).$$

	We note that the probability distributions of $\mathcal{A}$'s view and $\mathcal{S}_{\mathcal{A}}$'s output are identical.
	We thereby claim that Equation (1) holds.
	
	This completes the proof in the case of corrupted $\mathcal{A}$.
	
	\paragraph{Corrupted $\mathcal{B}$} In this case, we construct a probabilistic polynomial-time simulator $\mathcal{S}_{\mathcal{B}}$, when given the security parameter $\kappa$, $\mathcal{B}$'s input $sk_b$ and output $\langle\textbf{Z}\rangle_2$, can simulate the view of $\mathcal{B}$ in the protocol execution. To this end, we first analyze $\mathcal{B}$'s view $\mathsf{view}_{\mathcal{B}}^{\pi_2}([\![\textbf{Z}]\!]_b,sk_b,\kappa)$ in \textbf{Protocol 2}. The only message obtained by $\mathcal{B}$ is the ciphertext $[\![\langle\textbf{Z}\rangle_2]\!]_b$. Therefore, $\mathsf{view}_{\mathcal{B}}^{\pi_2}([\![\textbf{Z}]\!]_b,sk_b,\kappa)$ consists of $\mathcal{B}$'s input $sk_b$, the randomness $r_{\mathcal{B}}$, and the ciphertext $[\![\langle\textbf{Z}\rangle_2]\!]_b$.
	
	Given $\kappa$, $sk_b$, and $\langle\textbf{Z}\rangle_2$, $\mathcal{S}_{\mathcal{B}}$ generates a simulation of $\mathsf{view}_{\mathcal{B}}^{\pi_2}([\![\textbf{Z}]\!]_b,sk_b,\kappa)$ as follows. It encrypts $\langle\textbf{Z}\rangle_2$ with $\mathcal{B}$'s public key $pk_b$, and obtains $[\![\langle\textbf{Z}\rangle_2]\!]'_b$. Then, it generates $(sk_b,r_{\mathcal{B}},[\![\langle\textbf{Z}\rangle_2]\!]'_b)$ as the output.	Therefore, we have the following two equations:
	$$\mathsf{view}_{\mathcal{B}}^{\pi_2}([\![\textbf{Z}]\!]_b,sk_b,\kappa)=(sk_b,r_{\mathcal{B}},[\![\langle\textbf{Z}\rangle_2]\!]_b),$$
	$$\mathcal{S}_{\mathcal{B}}(1^\kappa, sk_b,\langle\textbf{Z}\rangle_2)=(sk_b,r_{\mathcal{B}},[\![\langle\textbf{Z}\rangle_2]\!]'_b).$$

	We note that both $[\![\langle\textbf{Z}\rangle_2]\!]_b$ and $[\![\langle\textbf{Z}\rangle_2]\!]'_b$ are the ciphertexts of $\langle\textbf{Z}\rangle_2$, and they look the same to $\mathcal{B}$. Therefore, the probability distributions of $\mathcal{B}$'s view and $\mathcal{S}_{\mathcal{B}}$'s output are identical.
	We thereby claim that Equation (2) holds.
	
	This completes the proof in the case of corrupted $\mathcal{B}$.

	In summary, \textbf{Protocol 2} securely computes $\mathcal{F}_{\mathsf{SSHEF}}$ in semi-honest model.

\end{proof}

\subsection{Secure Sparse Matrix Multiplication}

\begin{small}
\begin{framed}
\textbf{Functionality of Secure Sparse Matrix Multiplication $\mathcal{F}_{\mathsf{SSMM}}$}\\
    \\
    \textbf{Inputs:}
    \begin{itemize}
    	\item $\mathcal{A}$ inputs a sparse matrix $\textbf{X}$;
        \item $\mathcal{B}$ inputs a matrix $\textbf{Y}$.
    \end{itemize}
    \textbf{Outputs:}
    \begin{itemize}
        \item $\mathcal{A}$ outputs a share $\textbf{Z}_1$;
        \item $\mathcal{B}$ outputs a share $\textbf{Z}_2$, where $\textbf{Z}_1+\textbf{Z}_2=\textbf{X}\cdot\textbf{Y}$.
    \end{itemize}
\end{framed}
\end{small}

\paragraph{Security Proof} We show that \textbf{Protocol 1} is secure against semi-honest adversaries. Formally, we have the following theorem.

\nosection{Theorem 2}
\textit{
	Assume that the additively homomorphic crypto system $\Pi=(\textsf{KeyGen}, \textsf{Enc}, \textsf{Dec})$ is indistinguishable under chosen-plaintext attacks. Then, \textbf{Protocol 1} (denoted as $\pi_1$) is secure against semi-honest adversaries in $\mathcal{F}_{\mathsf{SSHEF}}$ model, as in \textbf{Definition 1}.
}

\begin{proof}
	We begin by proving the correctness of \textbf{Protocol 1}, \ie we prove that $\textbf{Z}_1+\textbf{Z}_2$ is equal to $\textbf{X}\cdot\textbf{Y}$.
	According to the protocol execution, $\mathcal{A}$ computes $[\![\textbf{Z}]\!]_b=\textbf{X}\cdot[\![\textbf{Y}]\!]_b$.
	From the additive homomorphism of the cryptosystem $\Pi$, we know that $[\![\textbf{Z}]\!]_b=[\![\textbf{X}\cdot\textbf{Y}]\!]_b$, \ie $\textbf{Z}=\textbf{X}\cdot\textbf{Y}$.
	After invoking $\mathcal{F}_{\mathsf{SSHEF}}$, $\mathcal{A}$ and $\mathcal{B}$ obtain the shares $\textbf{Z}_1$ and $\textbf{Z}_2$ satisfying $\textbf{Z}_1+\textbf{Z}_2=\textbf{Z}$.
	
	Therefore, it holds that $\textbf{Z}_1+\textbf{Z}_2=\textbf{X}\cdot\textbf{Y}$.	
	This proves the correctness of \textbf{Protocol 1}.
	
	We now prove that we can construct two simulators $\mathcal{S}_{\mathcal{A}}$ and $\mathcal{S}_{\mathcal{B}}$, such that
	\begin{eqnarray}
		\{\mathcal{S}_{\mathcal{A}}(1^\kappa,\textbf{X},\textbf{Z}_1)\}_{\textbf{X},\textbf{Y},\kappa}\cong\{\mathsf{view}_{\mathcal{A}}^{\pi_1}(\textbf{X},\textbf{Y},\kappa)\}_{\textbf{X},\textbf{Y},\kappa}, \\
		\{\mathcal{S}_{\mathcal{B}}(1^\kappa,\textbf{Y},\textbf{Z}_2)\}_{\textbf{X},\textbf{Y},\kappa}\cong\{\mathsf{view}_{\mathcal{B}}^{\pi_1}(\textbf{X},\textbf{Y},\kappa)\}_{\textbf{X},\textbf{Y},\kappa},
	\end{eqnarray}

	where $\mathsf{view}_{\mathcal{A}}^{\pi_1}$ and $\mathsf{view}_{\mathcal{B}}^{\pi_1}$ denotes the views of $\mathcal{A}$ and $\mathcal{B}$, respectively.
	
	We prove the above equations for a corrupted $\mathcal{A}$ and a corrupted $\mathcal{B}$, respectively.
	
	\paragraph{Corrupted $\mathcal{A}$} In this case, we construct a probabilistic polynomial-time simulator $\mathcal{S}_{\mathcal{A}}$ that, when given the security parameter $\kappa$, $\mathcal{A}$'s input $\textbf{X}$ and output $\textbf{Z}_1$, can simulate the view of $\mathcal{A}$ in the protocol execution. To this end, we first analyze $\mathcal{A}$'s view $\mathsf{view}_{\mathcal{A}}^{\pi_1}(\textbf{X},\textbf{Y},\kappa)$ in \textbf{Protocol 1}. In \textbf{Protocol 1}, the messages obtained by $\mathcal{A}$ are consisted of two parts. One is the ciphertext $[\![\textbf{Y}]\!]_b$; one is from the functionality $\mathcal{F}_{\mathsf{SSHEF}}$, \ie $\textbf{Z}_1$. Therefore, $\mathsf{view}_{\mathcal{A}}^{\pi_1}(\textbf{X},\textbf{Y},\kappa)$ consists of $\mathcal{A}$'s input $\textbf{X}$, the randomness $r_{\mathcal{A}}$, the ciphertext $[\![\textbf{Y}]\!]_b$, and $\textbf{Z}_1$.
	
	Given $\kappa$, $\textbf{X}$, and $\textbf{Z}_1$, $\mathcal{S}_{\mathcal{A}}$ generates a simulation of $\mathsf{view}_{\mathcal{A}}^{\pi_1}(\textbf{X},\textbf{Y},\kappa)$ as follows.
	
	\begin{itemize}
		\item $\mathcal{S}_{\mathcal{A}}$ randomly selects a matrix $\textbf{Y}'$, encrypts it with $pk_b$ and obtains $[\![\textbf{Y}']\!]_b$.
		\item $\mathcal{S}_{\mathcal{A}}$ simulates the functionality $\mathcal{F}_{\mathsf{SSHEF}}$ and takes $\textbf{Z}_1$ as the output for $\mathcal{A}$ in $\mathcal{F}_{\mathsf{SSHEF}}$.
		\item $\mathcal{S}_{\mathcal{A}}$ generates a simulation of $\mathsf{view}_{\mathcal{A}}^{\pi_1}(\textbf{X},\textbf{Y},\kappa)$ by outputting $(\textbf{X}, r_{\mathcal{A}}, [\![\textbf{Y}']\!]_b, \textbf{Z}_1)$.
	\end{itemize}
	
	Therefore, we have the following two equations:
	$$\mathsf{view}_{\mathcal{A}}^{\pi_1}(\textbf{X},\textbf{Y},\kappa)=(\textbf{X}, r_{\mathcal{A}},[\![\textbf{Y}]\!]_b, \textbf{Z}_1),$$
	$$\mathcal{S}_{\mathcal{A}}(1^\kappa,\textbf{X},\textbf{Z}_1)=(\textbf{X}, r_{\mathcal{A}},[\![\textbf{Y}']\!]_b, \textbf{Z}_1).$$

	We note that as the additively homomorphic cryptosystem $\Pi$ is indistinguishable under chosen-plaintext attacks, the probability distributions of $\mathcal{A}$'s view and $\mathcal{S}_{\mathcal{A}}$'s output are computationally indistinguishable.
	We thereby claim that Equation (3) holds.
	
	This completes the proof in the case of corrupted $\mathcal{A}$.
	
	\paragraph{Corrupted $\mathcal{B}$} In this case, we construct a probabilistic polynomial-time simulator $\mathcal{S}_{\mathcal{B}}$, when given the security parameter $\kappa$, $\mathcal{B}$'s input $\textbf{Y}$ and output $\textbf{Z}_2$, can simulate the view of $\mathcal{B}$ in the protocol execution. To this end, we first analyze $\mathcal{B}$'s view $\mathsf{view}_{\mathcal{B}}^{\pi_1}(\textbf{X},\textbf{Y},\kappa)$ in \textbf{Protocol 1}. The only message $\mathcal{B}$ receives is from the functionality $\mathcal{F}_{\mathsf{SSHEF}}$, \ie $\textbf{Z}_2$. Therefore, $\mathsf{view}_{\mathcal{B}}^{\pi_1}(\textbf{X},\textbf{Y},\kappa)$ consists of $\mathcal{B}$'s input $\textbf{Y}$, the randomness $r_{\mathcal{A}}$, and $\textbf{Z}_2$.
	
	Given $\kappa$, $\textbf{Y}$, and $\textbf{Z}_2$, $\mathcal{S}_{\mathcal{B}}$ simulates the functionality $\mathcal{F}_{\mathsf{SSHEF}}$ and takes $\textbf{Z}_2$ as the output for $\mathcal{B}$ in $\mathcal{F}_{\mathsf{SSHEF}}$.
	Then $\mathcal{S}_{\mathcal{B}}$ generates a simulation of $\mathsf{view}_{\mathcal{B}}^{\pi_1}(\textbf{X},\textbf{Y},\kappa)$ by simply outputting $(\textbf{Y},r_{\mathcal{A}},\textbf{Z}_2)$. Therefore, we have the following two equations:
	$$\mathsf{view}_{\mathcal{B}}^{\pi_1}(\textbf{X},\textbf{Y},\kappa)=(\textbf{Y}, r_{\mathcal{B}}, \textbf{Z}_2),$$
	$$\mathcal{S}_{\mathcal{B}}(1^\kappa,\textbf{X},\textbf{Z}_1)=(\textbf{Y}, r_{\mathcal{B}}, \textbf{Z}_2).$$

	We note that the probability distributions of $\mathcal{B}$'s view and $\mathcal{S}_{\mathcal{B}}$'s output are identical.
	We thereby claim that Equation (4) holds.
	
	This completes the proof in the case of corrupted $\mathcal{B}$.

	In summary, \textbf{Protocol 1} securely computes $\mathcal{F}_{\mathsf{SSMM}}$ against semi-honest adversaries in $\mathcal{F}_{\mathsf{SSHEF}}$ model.

\end{proof}

\end{document}